\documentclass[12pt]{article}

\usepackage{graphicx}
\usepackage{enumerate}
\usepackage{natbib}
\usepackage{url} % not crucial - just used below for the URL
\usepackage{amsfonts,amsmath,amsbsy,amsthm,amssymb,mathrsfs}
\usepackage{cases} %to number equations in big bracket.
\usepackage{anysize,indentfirst,geometry,xcolor}
\usepackage{authblk}

\usepackage{bm}
\usepackage{bigints} % big integral symbol
\usepackage{tikz}
\usepackage{caption}
\usepackage{subcaption}
\usepackage{multicol} % Used for the two-column layout of the document
\usepackage{multirow}% multirow allows you to combine rows in columns
\usepackage{setspace}

\theoremstyle{plain}
\newtheorem{thm}{Theorem}
\newtheorem{lem}{Lemma}
\newtheorem{cond}{Condition}
\newtheorem{rem}{Remark}

\def\bfX{\mathbf{X}}
\def\bfx{\mathbf{x}}

\def\bfalpha{\bm{\alpha}}
\def\hatalpha{\widehat{\bm{\alpha}}}
\def\bfgamma{\bm{\gamma}}
\def\bfomega{\bm{\omega}}

\def\tildee{\widetilde{e}}

\def\hatpsi{\widehat{\psi}}
\def\hateta{\widehat{\eta}}

\def\hatSigma{\widehat{\Sigma}}
\def\hatG{\widehat{G}}

\def\barS{\overline{S}}
\def\Pr{\operatorname{Pr}}

\title{Quantile Residual Lifetime Regression for Multivariate Failure Time Data}
\author{Tonghui Yu$^1$, Liming Xiang$^{1}$\thanks{Corresponding author. Email: LMXiang@ntu.edu.sg}, and Jong-Hyeon Jeong $^{2,3}$
\hspace{.2cm} \\ 
$^1$School of Physical and Mathematical Sciences, Nanyang Technological University, Singapore\\ 
$^{2}$Department of Biostatistics, Public Health, University of Pittsburgh, U.S.A.\\
$^{3}$Biometric Research Program,\\
Division of Cancer Treatment and Diagnosis\\
National Institutes of Health/National Cancer Institute, U.S.A.
}
\date{}
\begin{document}
\maketitle

\begin{abstract}
The quantile residual lifetime (QRL) regression is an attractive tool for assessing covariate effects on the distribution of residual life expectancy, which is often of interest in clinical studies.  When study subjects may experience multiple events of interest, the resulting failure times for the same subject are likely to be correlated. 
To accommodate such correlation in assessing the covariate effects on QRL, we propose a marginal semiparametric QRL regression model for multivariate failure time data. 
Our proposal facilitates parameter estimation using unbiased estimating equations, yielding estimators that are consistent and asymptotically normal. To address additional challenges in inference, we develop three approaches for variance estimation based on resampling techniques and a sandwich estimator, and further construct a Wald-type test statistic for hypothesis testing. The simulation studies and an application to real data offer evidence of the satisfactory performance and practical utility of the proposed method.studies and a real data analysis offer evidence of the satisfactory performance of the proposed method.
\end{abstract}


\doublespacing
\section{Introduction}

Multivariate failure times arise frequently in biomedical research when study subjects are exposed to multiple types of failure events, experience recurrent events in longitudinal studies, or are nested within clusters such as time to blindness in two eyes \citep{diabetic1976} and tooth extraction times \citep{caplan2005}.
Failure times obtained within the same cluster typically exhibit inherent association, which needs to be appropriately accounted for in the analysis of such data. 

Studying the distribution of residual lifetime generally provides valuable insights into disease prevention or treatment strategies for individuals at different life stages, especially for those who may not be at short-term risk of disease \citep{conner2022comparison}.
In the  Framingham heart study \citep{tsao2015}, each study subject may experience several cardiovascular diseases (events), such as coronary heart disease, myocardial infarction and hypertension, and potential dependence arises among the multiple disease event times obtained from a subject (cluster). It is interesting in this study to assess the effects of risk factors, e.g., BMI, blood pressure, cholesterol level, smoking and gender, on the distribution of remaining life times to the occurrence of each disease given that a subject is known to be disease-free at some followup time point. Since the dependence structure among multiple residual life times of a subject is unknown in practice, it poses both theoretical and computational challenges in regression analysis.   

Conventional methods for handling correlated failure times can be basically divided into three classes. The first explicitly models the dependence among multivariate failure times within a subject/cluster through frailty, which is often assumed to follow a known distribution from some positive scale family \citep{aalen1988, duchateau2008frailty}. The second employs copula functions to capture within-cluster association \citep{othus2010gaussian,kwon2022flexible,he2024analysis}.
The third, consisting of marginal models initially proposed by \cite{liang1986} for longitudinal outcomes, has been widely adopted and remains an active area of research. In particular, the marginal approach has been extensively studied in the context of multivariate survival data under 
the Cox proportional hazards and AFT models \citep [e.g.][]{cai1995estimating, jin2006rank, chen2010mul,lin2014marginal,  
xu2023marginal}, as well as censored quantile regression \citep {yin2005quantile, wang2009inference}. The basic idea of marginal models is to model the marginal distributions of multivariate outcomes as for independent observations, and treats associations among outcomes as a nuisance. Without specifying the correlation structure, this approach allows for more flexible, parsimonious models and is computationally more efficient than frailty or copula-based models. 

In this paper, we focus on the marginal method for regression analysis of multivariate residual lifetimes. 
As an alternative to conventional marginal models,
residual lifetime–based regression has attracted considerable attention in clinical studies due to its ease of understanding and capability to align with the demands in practice. For example, in cancer studies with patients who survived after some initial treatments, their remaining lifetimes are often of interest in evaluating the efficacy of the followup therapies. Compared to relative risks, the remaining life expectancy is more straightforward and readily understandable for patients.
Recently, the frailty model was extended to regression analysis of mean residual lifetimes in multicenter studies by \cite{huang2019} using a hierarchical likelihood approach. It is noted that failure times in biomedical studies often exhibit censorship, outliers and heteroscedasticity, which particularly leads to covariate effects on the remaining lifetimes varying over different follow-up stages. 
To this end,  quantile regression appears more appropriate than the mean-based regression for the remaining lifetimes. 

The quantile residual lifetime (QRL) regression, which leverages the strengths of censored quantile regression \citep{peng2008survival,wang2009locally},
examines the relationship between the quantile residual lifetimes and covariates and has gained growing attention recently. An overview of early developments can be found in the monograph by \cite{jeong2014statistical}. 
Semiparametric QRL regression analysis  has been investigated 
for univariate failure time outcomes. 
\cite{jung2009quantres} extended \cite{ying1995cqr}'s median regression model to quantile residual lifetimes and mimicked the least square estimating equations to construct an estimating equation for quantile coefficients. They suggested a grid search method to find some appropriate roots, 
which is computationally expensive especially in the presence of a large number of covariates because the estimating equation is neither monotone nor continuous. 
For testing significance, they studied a score-type test. 
\cite{zhou2011} and \cite{kim2012censored} proposed case-weighted empirical-likelihood ratio test.  Built upon \cite{jung2009quantres}'s method, \cite{ma2012} estimated quantile coefficient by spline smoothing instead and suggested a Wald-type test statistic. For data with longitudinal covariates, \cite{li2016quantile} and \cite{lin2019} developed an unbiased estimating equation that is solved via linear programming.
All these existing inferential methods for QRL models are under the independence assumption for failure times.

In the presence of multivariate or clustered failure times, applying these methods by ignoring possible correlations among outcome data may result in biases in variance estimation and loss of statistical power for testing hypotheses in consequence. To circumvent this issue, we study a marginal QRL regression model for multivariate failure time data, extending the idea of QRL regression \citep{li2016quantile} to accommodate the correlation among multiple failure time outcomes of a subject.  
We develop semiparametric estimating equations for parameter estimation and show theoretical properties of the resulting estimator regardless of the true dependence structures. A major hurdle in inference for QRL regression is variance estimation of parameter estimators. To this end,  we propose three methods to estimate the covariance matrix of the estimated regression coefficients accounting for the dependence of the multivariate failure times properly and compare their performance numerically.     

The rest of this article is organized as follows. In Section 2, we introduce notation for data and the proposed marginal QRL regression model first, and then provide the estimating equations for model parameters. In Section 3, we establish asymptotic properties of the resulting estimator and further develop variance estimation methods to facilitate inference. The performance of the proposed estimators is examined through extensive simulation studies in Section 4. We present an application to the analysis of the Flamingham Heart data in Section 5, followed by concluding remarks in Section 6. 

\section{Methodology} 

\subsection{Data and marginal QRL regression model} 

Consider a sample comprising $n$ clusters 
with each cluster containing $m_i$ observations. Consequently, the total number of observations in the sample amounts to $N = \sum_{i=1}^n m_i$.  Let $T_{ij}$ represent the $j$-th 
event time of cluster $i$ for $j=1,\ldots, m_i$  and $i=1,\ldots,n$,   and $\bfX_{ij}$ be the associated baseline covariate vector with the first element being one. 
At a specific time point $t_0$, we define $\theta_{\tau,t_0}$ as the $\tau$-th conditional quantile of the residual lifetime on a logarithmic scale, i.e., $\log(T_{ij}-t_0)$, conditional on the covariates $\bfX_{ij}$ and subject to the constraint $T_{ij}>t_0$. As a result, $\theta_{\tau,t_0}$ satisfies the equation $\Pr(\log(T_{ij}-t_0)\leq \theta_{\tau,t_0}|T_{ij}\geq t_0,\bfX_{ij}) = \tau$, which is equivalent to 
\begin{equation}
\Pr(t_0\leq T_{ij}\leq t_0+\exp(\theta_{\tau,t_0})|\bfX_{ij}) = \tau\Pr(T_{ij}\geq t_0|\bfX_{ij}). \label{qrl}
\end{equation}
For the $\tau$-th quantile of the remaining lifetimes among clusters whose event times are beyond time $t_0$, 
the linear QRL regression is assumed in the form of
\begin{equation}
\theta_{\tau,t_0} =  \bfX_{ij}^T\bfalpha_{\tau, t_0},
\label{eq:model0}
\end{equation}
where $\bfalpha_{\tau, t_0}$ is the vector of coefficients at time $t_0$ for covariate vector $\bfX_{ij}$ at some quantile level $\tau\in (0,1)$. Under model \eqref{eq:model0},
$T_{ij}$ can be modeled as
\begin{equation}
    \log (T_{ij}-t_0) = \bfX_{ij}^T\bfalpha_{\tau, t_0}+e_{ij}^{\tau}, j=1,\cdots, m_i, i=1,\cdots,n,
    \label{eq:model}
\end{equation}
where $e_{ij}^{\tau}$'s are correlated within the same  cluster but independent across clusters. 
For the sake of identifiability, we set the conditional $\tau$th quantile of $e_{ij}^{\tau}$ to zero given $\bfX_{ij}$ and $T_{ij}>t_0$.

\subsection{Estimation procedure} 
For ease of presentation, we omit $\tau$ and $t_0$ in the coefficient vector $\bfalpha$ in the following.  
When all survival times are exactly observed, the estimator of $\bfalpha$ can be obtained by solving the following estimating equations for $\bfalpha$:
\begin{equation}\label{esteq0}
\sum\limits_i\sum\limits_j\bfX_{ij}I(T_{ij}\geq t_0)[I\{T_{ij}\leq t_0+\exp(\bfX_{ij}^T\bfalpha)\}-\tau]=0.
\end{equation}

In the presence of right censoring, the survival outcome $T_{ij}$  is observed as $Y_{ij} = \min (T_{ij}, C_{ij})$ along with the censoring indicator 
$\Delta_{ij}=I(T_{ij} \leq C_{ij})$, 
and $C_{ij}$ is the corresponding censoring time.
It is assumed that $T_{ij}$ and $C_{ij}$ are 
independent, with $C_{ij}$
independently following a distribution characterized by the survival function $G(\cdot)$. 
With right-censored multivariate failure time data, one may modify equations in \eqref{esteq0} by adjusting censoring. Let $\bfalpha_0$ be the true value of $\bfalpha$. Note that 
\begin{equation}
\begin{split}
&E\left[\frac{\Delta_{ij}}{G(Y_{ij})}I\{Y_{ij}\leq t_0+\exp(\bfX_{ij}^T\bfalpha_0)\}\Bigg|Y_{ij}\geq t_0, \bfX_{ij}\right]  \\
&= \frac{\Pr\left\{t_0\leq T_{ij}=Y_{ij}\leq  C_{ij},T_{ij}\leq t_0+\exp(\bfX_{ij}^T\bfalpha_0)\Bigg| \bfX_{ij}\right\}}{\Pr\left\{C_{ij}\geq Y_{ij},Y_{ij}\geq t_0| \bfX_{ij}\right\}} \\
& = \frac{\Pr\left\{t_0\leq T_{ij}\leq t_0+\exp(\bfX_{ij}^T\bfalpha_0)| \bfX_{ij}\right\}}{\Pr\left\{T_{ij}\geq t_0| \bfX_{ij}\right\}G( t_0)} = \frac{\tau}{G( t_0)}. 
\end{split}
\end{equation}
 This motivates us to form a modified estimating equation for $\bfalpha$ as 
\begin{equation}
S_N(\bfalpha) = \frac{1}{N}\sum\limits_i\sum\limits_j \bfX_{ij}I(Y_{ij}\geq t_0)\left[\frac{\Delta_{ij}I\left\{Y_{ij}\leq t_0+\exp(\bfX_{ij}^T\bfalpha) \right\}}{\hatG(Y_{ij})/\hatG(t_0)} - \tau \right]=0,
\label{eq:estimating equation}
\end{equation}
where $\hatG(\cdot)$ is the Kaplan-Meier estimator of $G$ based on observations $\{Y_{ij},1-\Delta_{ij}\}$. The estimating equation (6) can be viewed as a multivariate version of \cite{li2016quantile}'s method, designed to account for the correlation among multivariate time-to-event data. In the independent cases with time-independent covariates, $S_N(\bfalpha)$ in \eqref{eq:estimating equation} 
reduces to the estimating function used in \cite{li2016quantile}. 
The estimator for $\bfalpha$, denoted by $\hatalpha$, can be obtained as
the solution to the estimating equations in \eqref{eq:estimating equation}. 
Equivalently, it is the minimizer of the following linear programming problem:
\begin{equation}
\begin{split}
\frac{1}{N}\sum\limits_i\sum\limits_j &
\frac{I(Y_{ij}\geq t_0)\Delta_{ij}}{\hatG(Y_{ij})/\hatG(t_0)} 
\rho_{\tau}\left\{\log(Y_{ij}-t_0)-\bfX_{ij}^T\bfalpha  \right\}\\
&+I(Y_{ij}\geq t_0)\rho_{\tau}\left\{A-\bfX_{ij}^T\bfalpha\left[1-\frac{\Delta_{ij}}{\hatG(Y_{ij})/\hatG(t_0)} \right]  \right\},
\label{eq:lp}
\end{split}
\end{equation}
where $\rho_{\tau}(u)=u[\tau-I(u<0)]$ is the quantile loss function, and
$A$ is a constant chosen to be exceptionally large such that 
$A> \max\limits_{i,j}\{\log(Y_{ij}-t_0)\}$. 
We adopt the fast interior point algorithm \citep{portnoy1997gaussian} to solve this linear programming problem, which can be readily implemented via function \emph{rq()} in R library \textbf{quantreg}  using the weighted QR model on the augmented data set comprising of pseudo responses $\{ (\log(Y_{11}-t_0),\cdots,\log(Y_{nm_n}-t_0),A,\cdots,A\}$ with corresponding covariates $\{\bfX_{11},\cdots,\bfX_{nm_n},\bfX_{11}^{*},\cdots,\cdots, \bfX_{nm_n}^{*}\}$ with $\bfX_{ij}^{*} = \left[1-\Delta_{ij}\hatG(t_0)/\hatG(Y_{ij}) \right]  \bfX_{ij}$. An alternative optimization algorithm analogue to \cite{li2015qr_semicompeting} may be considered, in which all artificial observations $\{\bfX_{ij}^{*}\}_{i,j}$ 
are treated as a whole unit. In some cases, these two competing optimization algorithms have negligible differences in parameter estimation when there is enough number of exactly observed residual lifetimes. For our motivating data, the optimization \eqref{eq:lp} produces much more reasonable results compared to results from a classical censored quantile regression (that is, $t_0=0$). This may be because \cite{li2015qr_semicompeting}'s method requires 
a large enough constant $A$ to bound the unified value $\sum\limits_i\sum\limits_j\left[1-\frac{\Delta_{ij}}{\hatG(Y_{ij})/\hatG(t_0)} \right]I(Y_{ij}\geq t_0)  \bfX_{ij}^T\bfalpha$ for any $\bfalpha$ in the parameter space, possibly leading to unstable estimation procedure  
especially when the magnitude of $\bfalpha$ or sample size is large.
 
\begin{rem}
{Once obtaining $\hatalpha$, the $\tau$-th conditional quantile of the logarithm of the residual lifetime for a specific individual with covariates $\bfx$ can be estimated as $\widehat{\theta}_{\tau,t_0}=\bfx^T\hatalpha_{\tau, t_0}$. In practice,  the estimated conditional quantiles $\widehat{\theta}_{\tau,t_0}$ may not be monotonically increasing in $\tau$ due to lack of sufficient data and/or quantile crossing. To account for the nonmonotonicity problem,
one may follow the rearrangement method developed by \cite{chernozhukov2010cross} to construct monotone quantile curves by using the order statistics of the rearranged quantile estimates. The rearrangement procedure 
has been commonly used in various quantile regression models \citep{wang2012estimation,wang2013estimation,yu2021}. 
Given asymptotic properties of $\hatalpha$ and the Wald-type inference discussed in Section 3, the confidence intervals for $\widehat{\theta}_{\tau,t_0}$ can be subsequently constructed. As shown by \cite{chernozhukov2010cross} and \cite{wang2012estimation} either theoretically or numerically, 
the quantile estimators with/without rearrangement exhibited nearly identical performance in estimation and inference.}
\end{rem}

\section{Asymptotic Properties and Inference}
\subsection{Consistency and asymptotic normality}

The following conditions are necessary to derive the asymptotic properties of the proposed estimator 
obtained from solving the estimating equation in \eqref{eq:estimating equation}. To associate with the total sample size $N$, in this section, we rewrite $\hatalpha$  as $\hatalpha_N$.
\begin{cond}
The parameter space $\mathcal{D}$ for $\bfalpha$ is a compact region with $\bfalpha_0$ in the interior. For any $\bfalpha\in\mathcal{D}$, there exists $t_u$ such that $\Pr\left\{ \log(Y_{ij}-t_0)\geq t_u)|\bfX_{ij}\right\} $ is uniformly bounded away from zero and $\bfX_{ij}^T\bfalpha\leq t_u$ with probability one.
\label{cond:ProbY}
\end{cond}
\begin{cond}
The estimator $\hatG$ has $\sup\limits_{t\leq t_u}|\hatG(t) - G(t)| = o(N^{-1/2+\epsilon})$ for any $\epsilon>0$.
\label{cond:Gest}
\end{cond}

\vspace{-1.1cm}
\begin{cond}
Given $T_{ij}>t_0$, the conditional distribution functions $F_e(e|\bfX_{ij}) =\Pr(e_{ij}^{\tau}\leq e|\bfX_{ij})$ have densities  $f_e(\cdot|\bfX_{ij})$ which is Lipschitz continuous in the neighborhood of 0. We assume that
$N^{-1}\sum_{ij}\Pr(T_{ij}\geq t_0|\bfX_{ij})f_e(0|\bfX_{ij})\bfX_{ij}\bfX_{ij}^T$ converges almost surely to a positive definite and bounded matrix, denoted by $\Lambda$.
\label{cond:f_e}
\end{cond}

\begin{cond}
1) The cluster size $m_i$ is finite. 2) $\bfX_{ij}$ are uniformly bounded for $j=1,\cdots,m_i$ and $i=1,\cdots,n$.
\label{cond:m_i}
\end{cond}

Condition \ref{cond:ProbY} is an standard condition for quantile residual lifetime regression and commonly imposed in literature \citep{jung2009quantres}. Condition \ref{cond:Gest} holds in most cases when $\hat G$ is the Kaplan-Meier estimator \citep{csorgHo1983rate}.   Condition \ref{cond:f_e} is to ensure $\inf\limits_{\bfalpha: ||\bfalpha - \bfalpha_0|| = \epsilon} ||\barS_N(\bfalpha)||> 0 $ for any $\epsilon>0$, which is needed to establish consistency of the estimator $\hatalpha_N$.  Matrix $\Lambda$, defined in Condition \ref{cond:f_e}, will be part of the slope matrix in estimator' asymptotic variance, and its positive definiteness and boundness guarantee the asymptotic normality of the estimator $\hatalpha_N$.  Condition \ref{cond:m_i} is a weak assumption and commonly seen in literature.

To justify the asymptotic properties of the proposed estimator, we consider the conditional expectation of the proposed estimating function \eqref{eq:estimating equation} with substitution of true censoring distribution and define
\begin{equation}
 %\begin{split}
\barS_N(\bfalpha)
= \frac{1}{N}\sum\limits_i\sum\limits_j \bfX_{ij}G(t_0)\left[\Pr\left\{ t_0\leq T_{ij}\leq t_0 + \exp(\bfX_{ij}^T\bfalpha)|\bfX_{ij}\right\} - \tau \Pr\left\{T_{ij}\geq t_0|\bfX_{ij}\right\}\right].
% \end{split}
\end{equation}
By the definition of quantile residual lifetime function in equation \eqref{qrl}, it follows that $\bfalpha_0$ is the unique root of $\barS_N(\bfalpha) = 0$ for some commonly used distributions of the failure time $T$, provided that its survival function is continuous and strictly decreasing with a closed form \citep{jeong2014statistical}. 
\begin{thm}
(Consistency.) Under Conditions \ref{cond:ProbY}-\ref{cond:f_e}, $\hatalpha_N$ satisfying $S_N(\hatalpha_N) = o(1)$ converges almost surely to $\bfalpha_0$ as $N\rightarrow\infty$. 
\end{thm}

\begin{lem}
If Condition \ref{cond:ProbY} holds, $N^{1/2}S_N(\bfalpha_0)$ converges to a zero-mean normal distribution 
with the asymptotic covariance matrix $\Sigma$ as defined in the proof of this lemma in the Appendix.
\label{lem:s_n}
\end{lem}

\begin{thm}
(Asymptotic normality.)  Under Conditions \ref{cond:ProbY}- \ref{cond:m_i}, $\hatalpha_N$ satisfying $S_N(\hatalpha_N) = o(N^{-1/2})$ is asymptotically normal, i.e.,
$N^{1/2}(\hatalpha_N-\bfalpha_0)\xrightarrow{d} N(0,V(\bfalpha_0))$, where $V(\bfalpha_0) = \widetilde{\Lambda}^{-1}\Sigma\widetilde{\Lambda}^{-1}$, $\widetilde{\Lambda}= G(t_0)\Lambda$.
\end{thm}
The proofs of Lemma 1 and Theorems 1-2 are provided in the Appendix. The main challenge in proving asymptotic properties is to account for association among multivariate failure time data. To our knowledge this issue has not been addressed in the literature regarding parameter estimation in quantile residual lifetime regression.
To prove the consistency of the proposed estimator, we adopt arguments established by \cite{resnick2019prob} in the L\'{e}vy's theorem, \cite{white1980nonlinear} in their Lemma 2.2 and \cite{van2000asymptotic} in Theorem 5.9. 
Based on martingale processes involving in estimation of censoring distribution as well as the Lyapunov central limit theorem, we can show 
results in our new Lemma 1. The asymptotic normality of $\hatalpha_N$ follows from Lemma 1 and similar arguments developed by \cite{he1996Bahadur} and \cite{wang2009inference}. It is worth noting that Condition \ref{cond:m_i} is essential for applying \cite{he1996Bahadur}'s theorems to the model we considered. In fact, this condition can be extended to the situation in which $m_i = o(n^{\varrho})$ holds for some constant $0<\varrho<1/5$ and $N^{-1}\sum\limits_{i}\sum\limits_{j,k}\bfX_{ij}\bfX_{ik}^T <\infty$.

\begin{rem}
Though our estimating function for the regression coefficients essentially keeps the same form as that for univariate survival data given by \cite{li2016quantile}, the asymptotic variances of the estimated regression coefficients address the association among multivariate data. To further illustrate this, we consider the error terms in model (3) having an exchangeable correlation structure as an example.
Suppose that $\Pr(e_{ij}\leq 0, e_{ij'}\leq 0|T_{ij}>t_0,T_{ij'}>t_0,\bfX_{ij},\bfX_{ij'}) =\delta$ for any $j\neq j'$, where $\delta$ measures the within-cluster dependence.
In this case, the middle matrix $\Sigma$ in variance matrix $\bm{V}(\alpha_0)$ is the sum of the following three components: $I_1 = \frac{1}{N}\sum\limits_{i=1}^n \operatorname{Var}\psi_i(\bfalpha_0)$, $I_2 = \frac{1}{N}\sum\limits_{i=1}^n \operatorname{Var}\eta_i(\bfalpha_0)$, $I_3 = \frac{2}{N}\sum\limits_{i=1}^n \operatorname{Cov}(\psi_i(\bfalpha_0),\eta_i(\bfalpha_0))$, where $\psi_i$ and $\eta_i$ are defined in the supplementary material. After some calculations, $I_1$ can be written as  
\begin{equation*}
\begin{split}
I_1 &= \frac{1}{N}\sum\limits_{i=1}^n \sum_{j} \bfX_{ij}\bfX_{ij}^TI(Y_{ij}\geq t_0)\left(\frac{\tau G(t_0)}{G(Y_{ij})}-\tau^2\right) \\
&+\frac{1}{N}\sum\limits_{i=1}^n \sum_{j,j'}\bfX_{ij}\bfX_{ij'}^TI(Y_{ij}\geq t_0,Y_{ij'}\geq t_0)\left(\delta-\tau^2\right).
\end{split}
\end{equation*}
The explicit expressions for $I_2$ and $I_3$ are  complicated and lengthy, and thus omitted here for brevity.  It is noted that when $\delta\in (\tau^2,\tau]$, the errors are positively correlated, whereas for $\delta\in[0,\tau^2)$, they are negatively correlated. When $\delta = \tau^2$, the errors are independent. Ignoring the within-cluster dependence by assuming $\delta = \tau^2$ leads to biased estimation for the asymptotic standard deviation of the estimator $\widehat{\alpha}_N$.
\end{rem}

\subsection{Inference} 

The asymptotic normality of the proposed estimator established in Theorem 2 offers evidence for the feasibility of the Wald-type inference and construction of confidence intervals. An additional challenge for inference is to estimate the variance of the proposed estimator. 
To directly estimate the asymptotic variance matrix $V(\bfalpha_0)$ is impractical since it takes a complicated form involving the unknown error density function $f_e(0|\bfX_{ij})$ for computing $\Lambda$ and unknown censoring distribution function in $\Sigma$. 
To overcome this problem, we develop 
three approaches, including a perturbation resampling, sandwich estimators and  multiplier bootstrap based sandwich estimators, for asymptotic variance estimation of $\hatalpha_N$. 

\subsubsection{Resampling method}

It is worthwhile noted that the conventional %familar pairwise
bootstrapping by sampling with replacement is not appropriate
for data from the longitudinal/clustered studies. 
To this end, a feasible way is to bootstrap and repeatedly solve a perturbation version of \eqref{eq:estimating equation}.
\cite{jin2003rank} proposed analogous perturbation resampling procedure for estimating the limiting variance matrices in AFT models without requiring density estimation or numerical derivatives and showed its validity. This resampling approach has been widely applied to survival data especially when estimating equations are non-smooth \citep{yin2005quantile,peng2008survival,li2016quantile}. 
It is also applicable for a wide variety of models and not limited to independent cases \citep{hagemann2017cluster,galvao2023bootstrap}.
To obtain a consistent variance estimator, we consider a similar perturbation resampling method and account for the possible heterogeneity in the data. 

In particular, we first generate independent and identically distributed positive multipliers $\gamma_i$ from an exponential distribution with $E(\gamma_i) = \operatorname{Var}(\gamma_i)=1$, for  $i=1,..,n$. 
We define the randomly perturbed version of $S_N(\bfalpha)$ as
\begin{equation}
S_N^{*}(\bfalpha) = \frac{1}{N}\sum\limits_i\gamma_i\sum\limits_j \bfX_{ij}I(Y_{ij}\geq t_0)\left[\frac{\Delta_{ij}I\left\{Y_{ij}\leq t_0+\exp(\bfX_{ij}^T\bfalpha) \right\}}{G^{*}(Y_{ij})/G^{*}(t_0)} - \tau \right]
\label{eq:perturbed estimating equation}
\end{equation}
and $G^{*}(\cdot)$ in \eqref{eq:perturbed estimating equation} is a perturbed version of the Kaplean-Meier estimator 
in the form of 
\begin{equation*}
G^{*}(t) = \prod\limits_{ij: Y_{ij}\leq t} \left\{ 1-\frac{d \overline{N}^{*}(Y_{ij})}{\overline{Y}^{*}(Y_{ij})} \right\},
\end{equation*}
where $\overline{N}^{*}(t) = \sum\limits_{k=1}^{n}\gamma_k\sum\limits_{l=1}^{m_i} I(\delta_{kl}=0,Y_{kl}\leq t)$, $\overline{Y}^{*}(t) = \sum\limits_{k=1}^{n}\gamma_k\sum\limits_{l=1}^{m_i} I(Y_{kl}\geq t)$ and $d \overline{N}^{*}(t) = \overline{N}^{*}(t)-\overline{N}^{*}(t-)$. 
Then the resampled estimate $\hatalpha^{*}$ is obtained by solving the updated estimating equations 
$S_N^{*}(\bfalpha) =0$. 

\begin{thm}
Under Conditions \ref{cond:ProbY}- \ref{cond:m_i}, the conditional distribution of $N^{1/2}(\hatalpha^{*}-\hatalpha)$ given the observed data converges to the same limiting distribution of $N^{1/2}(\hatalpha_N-\alpha_0)$. 
\end{thm}

The proof of Theorem 3 is in line with that of Theorem 2 to some extent, with its sketch given in the Appendix.
Therefore, the variance of $\hatalpha_N$ can be estimated using 
the sample variance of $B$ 
resampled estimates, $(\hatalpha^{*(1)},\cdots,\hatalpha^{*(B)})$, which are obtained by
repeating the above resampling procedure for $B$ times. 

\subsubsection{A closed-form sandwich estimator }

We consider estimation of  $\Lambda$ first, in which  
$f_e(0|\bfX_{ij})$ is unknown. \cite{wang2019copula} proposed 
quantile regression with correlated data and estimate $f_e(0|\bfX_{ij})$  by a well-known quotient estimation method. Specifically, based on the large-sample behavior of regression quantile spacing shown by \cite{goh2009nonstandard}, the conditional density function $f_e(0|\bfX_{ij})$ can be consistently estimated using the difference quotient  
\begin{equation}
    \hat{f}_e(0|\bfX_{ij}) = \frac{2h_N}{\bfX_{ij}^T\left[\hatalpha_{\tau+h_N,t_0}-\hatalpha_{\tau-h_N,t_0}\right]},
\label{eq:fe_est}
\end{equation}
where $h_N$ is a bandwidth parameter such that $h_N\rightarrow 0$ as $N$ goes to infinity,  $\hatalpha_{\tau+h_N,t_0}$ and $\hatalpha_{\tau-h_N,t_0}$ are the roots of the estimating equation in \eqref{eq:estimating equation} at the residual time point $t_0$ and two specific quantile levels $\tau+h_N$ and $\tau-h_N$. In practice, we follow \cite{hall1988distribution} and take $$h_N = 1.57N^{-1/3}\left[1.5\phi^2\left\{\Phi^{-1}(\tau) \right\}/(2\left\{ \Phi^{-1}(\tau)\right\}^2+1) \right]^{1/3},$$ where $\phi$ and $\Phi$ are the density and distribution functions of the standard normal distribution, respectively. It follows from \cite{hendricks1992hierarchical} that 
$\widehat{\widetilde{\Lambda}} =N^{-1}\sum_{ij}\hatG(t_0)I(Y_{ij}\geq t_0)\hat{f}_e(0|\bfX_{ij})\bfX_{ij}\bfX_{ij}^T\xrightarrow{P} \widetilde{\Lambda}$.
As pointed by \cite{koenker2005quantile}, the crossing issues may occur in the estimated conditional quantile planes, and so 
for implementation we may
replace $\hat{f}_e(0|\bfX_{ij})$ simply by its positive part in \eqref{eq:fe_est}, that is,
\begin{equation}
    \hat{f}_e(0|\bfX_{ij}) = \max\left\{0,\frac{2h_N}{\bfX_{ij}^T\left[\hatalpha_{\tau+h_N,t_0}-\hatalpha_{\tau-h_N,t_0}\right]-\epsilon}\right\},
\label{eq:fe_est2}
\end{equation}
where $\epsilon$ is a small positive constant used to 
avoid dividing by zero in some rare cases.
In addition to equations \eqref{eq:fe_est}-\eqref{eq:fe_est2}, another approach based on a kernel density estimation  
proposed by \cite{powell1991estimation} can be adopted for estimating $f_e(0|\bfX_{ij})$ and the consistency of $\widehat{\widetilde{\Lambda}}$ maintains as well \citep{koenker2005quantile}. 

To estimate $\Sigma$, we can use a closed-form direct approximation given by
\begin{equation*}
\begin{split}
\widehat{\Sigma}&= N^{-1}\sum_{i=1}^n (\hatpsi_i+ \hateta_i)(\hatpsi_i+ \hateta_i)^T,
\end{split}
\end{equation*}
where
\begin{equation*}
\hatpsi_i= \sum\limits_{j=1}^{m_i} \bfX_{ij}I(Y_{ij}\geq t_0)\left[\frac{\Delta_{ij}I\left\{Y_{ij}\leq t_0+\exp(\bfX_{ij}^T\hatalpha_N) \right\}}{\hatG(Y_{ij})/\hatG(t_0)} - \tau \right],
\end{equation*}
\begin{equation*}
\begin{split}
\hateta_i
&=  \sum\limits_{k,j} \frac{\bfX_{kj}\delta_{kj}I\left\{t_0\leq Y_{kj}\leq t_0+\exp(\bfX_{kj}^T\hatalpha_N) \right\}}{\hatG(Y_{kj})/\hatG(t_0)}\times\\
&\qquad\left[\sum\limits_{l}\dfrac{(1-\delta_{il})I(t_0\leq Y_{il}\leq Y_{kj})}{\sum_{r,s}I(Y_{rs}\ge Y_{il})}- \sum\limits_{l}\sum\limits_{u,v}\dfrac{(1-\delta_{uv})I\{Y_{uv}\leq\min(Y_{il},Y_{kj})\}}{(\sum_{r,s}I(Y_{rs}\ge Y_{uv}))^2}\right]
\end{split}
\end{equation*}
with plugging in the Kaplan-Meier estimator $\hat G(\cdot)$ of $G(\cdot)$. This direct approximation replaces all unknown quantities in $\Sigma$ with their sample version estimates, which have closed forms but are complicated 
owing to the martingale processes and non-parametric estimation for cumulative hazard function of the censoring variable. 

Based on the sandwich estimator with estimated slope matrix 
$\widehat{\widetilde{\Lambda}}$ and estimated middle matrix 
$\widehat{\Sigma}$, the Wald-type statistic for testing the hypothesis $H_0 :\bfalpha = \bfalpha_0$ can be consequently constructed by 
$\mathcal{W}_N = N(\hatalpha_N-\bfalpha_0)^T(\widehat{\widetilde{\Lambda}}^{-1}\hatSigma\widehat{\widetilde{\Lambda}}^{-1})^{-1}(\hatalpha_N-\bfalpha_0)$. It follows from 
the Slutsky's theorem that $\mathcal{W}_N$ converges in distribution to the same limiting distribution as $N(\hatalpha_N-\bfalpha_0)^T(\widetilde{\Lambda}^{-1}\Sigma\widetilde{\Lambda}^{-1})^{-1}(\hatalpha_N-\bfalpha_0)$. Then the  conventional $\chi^2$ test can be applied to test the hypothesis about regression coefficients at some specific quantile level. 

\subsubsection{Resampling-based Sandwich estimator} 

By integrating the strengths of both the resampling and sandwich estimator approaches, we develop a resampling-based sandwich estimator to improve the accuracy of the sandwich estimator, while circumventing repeatedly solving equation \eqref{eq:perturbed estimating equation} in the resampling method. Similar methods have been studied by 
\cite{zeng2008resampling} and \cite{chiou2015aft} under different models to achieve
consistent variance estimators.

For the asymptotic variance matrix $V(\bfalpha_0) = \widetilde{\Lambda}^{-1}\Sigma\widetilde{\Lambda}^{-1}$, estimators of $\widetilde{\Lambda}$ and $\Sigma$ can be obtained from a computationally efficient resampling procedure without the need of solving estimating equations. Given a set of random multipliers $(\gamma_1,\cdots,\gamma_n)$ generated as in Subsection 3.2.1, the perturbed estimating function  
$S_N^{*}(\bfalpha)$ in \eqref{eq:perturbed estimating equation} evaluated at the estimate $\hatalpha_N$ (the root of equations in \eqref{eq:estimating equation}) is obtained. Then repeating this $B$ times, we obtain the set $\{S_N^{*(k)}(\hat{\bfalpha}_N), k=1,\cdots, B\}$, and the sample variance of $\{\sqrt{N}S_N^{*(k)}(\hat{\bfalpha}_N), k=1,\cdots, B\}$ provides the resampled estimate of $\Sigma$, denoted by $\hatSigma^*$. 
Next, we generate $B$ random samples, denoted by $\{Z_k, k=1,\cdots,B\}$, from a multivariate normal distribution with mean zero and covariance matrix $(\hatSigma^*)^{-1}$.
Following the resampling method given by \cite{zeng2008resampling}, the inverse of the sample covariance matrix of $\{\sqrt{N}S_N(\hatalpha_N+ N^{-1/2}Z_k), k=1,\cdots, B \}$ can be used as a consistent estimator of $V(\bfalpha_0)$. 

\section{Simulation studies}
In this section, we conduct 
simulation studies to assess the performance of the proposed estimators in various situations. Particularly, data are generated with individual-level covariates in Scenarios 1, 
with cluster-level covariates and different marginal distributions of error terms in Scenarios 2-3, 
and with heterogeneous errors in Scenario 4. 
We also examine the performance of the proposed methods under various types of dependence structures in Scenarios 5-7, 
and for the case with multiple covariates in Scenario 8. 
The simulation setups for Scenarios 1-4 are provided below, while those for Scenarios 5-8 are detailed in the Supplementary Material.

\textbf{Scenario 1} 
is designed to evaluate the finite sample performance of our proposal under  the longitudinal study with an individual-level covariate.
For each observation case $j$ of individual $i$ with $j=1,\cdots, m$ and $i=1,\cdots, n$, 
we generate a single baseline covariate, $x_{ij}$, independently from a uniform distribution %spanning
on the interval $[0,1]$, and survival outcome $T_{ij}$ following a multivariate accelerated failure time model in the form of
\begin{equation}
    \log T_{ij} = \beta_0+\beta_1 x_{ij}+ \epsilon_{ij},
    \label{ex:model}
\end{equation}
%Here, 
where $\exp(\epsilon_{ij})$ marginally follows an exponential distribution with the rate parameter %of 
$\lambda=0.69$. We construct the joint distribution of $(\epsilon_{i1},\cdots,\epsilon_{im})$ through 
a Clayton copula with Kendall's tau of 0, 0.5 and 0.8, corresponding to the independent, moderate correlated and strongly correlated cases, respectively. 
We take the values of $(\beta_0,\beta_1)$ as $(1,1)$. 
Under model \eqref{ex:model} with the above setting, parameters in the corresponding  quantile residual lifetime model \eqref{eq:model} are given by 
$\bfalpha_0(\tau,t_0) = (\alpha_0(\tau,t_0),\alpha_1(\tau,t_0))$, % with 
where $\alpha_0(\tau,t_0) = \log[-\lambda^{-1}\log(1-\tau)]+\beta_0$ and $\alpha_1(\tau,t_0) = \beta_1$. 

In \textbf{Scenario 2,} 
a cluster-level covariate is considered in the working AFT model \eqref{ex:model} to mimic community randomized studies or patients with multiple disease progressions in practice. The scheme for data generation is same 
as in Scenario 1 except taking  $x_{ij}=x_{i}$ with $x_{i}$ being generated from $\text{Uniform}(0,1)$ for all  $j=1,\cdots m$ observations of cluster $i$.
A Clayton copula joint distribution is also considered for the error terms with  Kendall's tau equal 0.5. 

\textbf{Scenario 3} 
considers a residual lifetime model with error term marginally from a logistic distribution. Same as in Scenario 2, a cluster-level covariate $x_{i}$ is independently from $\text{Uniform}[0,1]$. The failure time outcome $T_{ij}$ is generated from the residual lifetime model:
\begin{equation}
    \log (T_{ij}-t_m) = \beta_0+\beta_1 x_{i}+ \sigma\epsilon_{ij},
    \label{ex:model3}
\end{equation}
where $\epsilon_{ij}$ marginally follows a standard logistic distribution, leading to a baseline log-logistic distribution for the residual lifetime $T_{ij}-t_m$. The joint distribution of $(\epsilon_{i1},\cdots,\epsilon_{im})$ %follows 
is given by a Clayton copula % structure, 
with Kendall's tau equal %set to 
0.5.
We take $t_m=1$, $\beta_0=1$, $\beta_1=0$ and $\sigma=0.5$ 
in \eqref{ex:model3}, %which suggests that 
corresponding to $\alpha_1(\tau,t_0) = 0$ and 
$$\alpha_0(\tau,t_0) = \left\{\begin{array}{cc}
  \log\left[ \exp\left(\sigma\log\left(\frac{\tau}{1-\tau}\right)+\beta_0\right)-t_0+t_m\right],   &  t_0\leq t_m\\
  \log\left[ \left(\frac{\tau+ \exp(-\frac{\beta_0}{\sigma})(t_0-t_m)^{\frac{1}{\sigma}}}{1-\tau}\right)^{\sigma}\exp(\beta_0)-t_0+t_m\right],   &  t_0>t_m
\end{array}\right.$$
in model \eqref{eq:model}.

\textbf{Scenario 4} is designed to illustrate the substantial gains of the quantile residual lifetime regression compared to the Cox or AFT model, particularly in handling heterogeneous data and revealing how covariate effects vary across different quantile levels. The failure time outcome $T_{ij}$ follows the model given by $\log T_{ij} = \beta_0+\beta_1x_{ij} +(1-a x_{ij})\epsilon_{ij}$, where  $\beta_0=1$, $\beta_1 = 2$, covariate $x_{ij}=x_{i}$ and $x_i\sim \text{Bernoulli}(0.5)$. The degree of heteroscedasticity rises with increasing values of $a$. The generation of $(\epsilon_{i1},\cdots,\epsilon_{im})$ is the same as in Scenario 2 except the rate parameter $\lambda=2$. Consequently, the true regression coefficients in model (3) are $\alpha_0(\tau,t_0) = \log[-\lambda^{-1}\log(1-\tau)]+\beta_0$ and 
\begin{equation*}
\alpha_1(\tau,t_0) =\left\{\begin{array}{cc}
  -a\log(-\lambda^{-1}\log(1-\tau))+\beta_1,   &  t_0=0,\\
   \log\left\{\frac{t_0[1-\lambda^{-1} t_0^{-1/(1-a)}\log(1-\tau)\exp(\frac{\beta_0+\beta_1}{1-a})]^{1-a}-t_0}{-\lambda^{-1}\log(1-\tau)\exp(\beta_0)} \right\},  & t_0\neq 0.
\end{array}\right.
\end{equation*}
Their values, determined for 
$\tau=0.25, 0.5$ and $t_0=0,1,2$, can be found in Table \ref{table: sim4}. 
Under this setup, with a fixed quantile level, the covariate effect decreases as $t_0$ increases. While given a fixed $t_0$, the covariate effect decreases as the quantile level increases. 

In all scenarios, we consider the cluster size 
$m=3$ or 10. The number of clusters is configured as $n=200$ or $500$. The censoring time variable $C_{ij}$ is generated from a uniform distribution over the interval $[0,20]$, achieving a censoring rate between $20\%$ and $40\%$. 

Based on 500 simulated data sets for each simulation setting, 
results of the estimation of regression coefficients $\alpha_0$ and $\alpha_1$ are summarized in Tables \ref{table: sim1-2}-\ref{table: sim4} for Scenarios 1-4 and Tables S.1-S.4 for Scenarios 5-8 in the Supplementary Material, respectively, in terms of averaged bias of point estimates, the Monte Carlo standard derivation (MCSD) of point estimates, the average of standard error (ASE), and the empirical coverage percentage (CP) of the $95\%$ confidence intervals. 
For standard errors, we report the results 
of three variance estimators: the fully resampling method (FR), the closed-form sandwich estimator (CFS) and the resampling-based sandwich estimator (RBS) proposed in subsection 3.2, in comparison with the fully resampling estimator of variance proposed by \cite{li2016quantile} for independent failure times (IFR) with time-independent covariates. All perturbation resampling-based estimators of variance are computed based on $B=500$ multiplier replicates. 

In general, it can be seen from these tables that the estimated regression coefficients appear to be asymptotically unbiased. Biases and standard deviations of point estimates decrease as the number of clusters or cluster size increases. 
Their standard errors obtained from FR/CFS/RBS are generally close to the Monte Carlo empirical standard deviation of the estimates. The coverage probabilities based on FR/CFS/RBS variance estimators also reasonably approach the nominal level 0.95. 

To be specific, as shown in Supplementary Table S.1 under Scenario 1 with independent survival outcomes, %\ref{table: sim1-1}  
IFR, FR, CFS and RBS variance estimators yield similar results in terms of average estimated standard error as well as coverage probability. With stronger
dependence %If we increase associationamong survival 
among failure time outcomes, corresponding to higher values of Kendall's tau %within the same clusters, 
as demonstrated in Table \ref{table: sim1-2} and Table S.2, %\ref{table: sim1-3}, 
the IFR estimator is more likely to underestimate %causes underestimation for 
standard errors particularly for $\alpha_0$ as the cluster size increases. Similar trends can be found across various situations in Scenarios 2-8. As shown in Tables \ref{table: sim2}-\ref{table: sim4} and Tables S.3-S.4, the IFR estimator generally yields considerably lower ASEs than the benchmark MCSDs in most cases, along with the empirical CPs below the nominal 95\%. Such an issue becomes more pronounced--iparticularly for coefficients associated with cluster-level covariates--as the correlation among failure times strengthens, the cluster size grows, or $t_0$ is small. 
This underperformance is mainly attributed to the fact that the IFR method utilizes a conventional resampling approach, which treats the data $\{(Y_{ij},\delta_{ij}, \bfX_{ij})\}$ as if they are independent 
and samples from them with replacement across all $(i,j)$, thereby ignoring the correlation among multivariate
failure times. 
It is observed that the performance of the IFR estimator improves as  $t_0$  increases, especially when $t_0=2$ in our simulation setups. A possible reason for this improvement is that both the values of MCSD and ASE increase as a result of smaller sample sizes under the restricted population where $T_{ij}> t_0$. 

On the other hand, the proposed estimators closely match MCSD and produce reasonable coverage probabilities near the nominal level, highlighting the importance of accounting for within-cluster dependence to ensure accurate variance estimation and reliable inference. Results summarized in Supplementary Tables S.3-S.4 demonstrate the outperformance of the proposed marginal method in accommodating diverse dependence structures for multivariate failure times even though an independent working model is used. These promising findings further exhibit a degree of robustness of the method across different types of copula.

Each of the three variance estimators we proposed has unique advantages. The FR estimator provides the best performance but is less computationally efficient, requiring at least 55\% more time than the RBS estimator. The CFS estimator stands out for its elegant form and computational efficiency. However, the CFS estimator tends to be slightly conservative with higher CP for larger $t_0$, possibly due to bandwidth selection, and becomes inestimable at high quantile level as indicated by the difference quotient in Equation \eqref{eq:fe_est}. The RBS estimator is a trade-off between computational efficiency and accuracy, performing well in most scenarios. While it falls behind the FR estimator in a few cases, its performance improves with larger sample sizes, making it the most practical choice.

As a final remark, it is worth noting that when $t_0$ and the quantile level $\tau$ are large, the number of individuals with exactly observed failure times would become limited, leading to potential identifiability issues. 
To ensure identifiability, we therefore consider estimation at $\tau=0.5$ under various scenarios restricting the censoring rate to below 50\% in the simulation studies. When using a quantile level $\tau\in(0,0.4)$ and a higher censoring rate, e.g., 62\% as in the following real data analysis, the proposed estimator exhibits similar performance. Thus, the corresponding simulation results are omitted.

\section{An illustrative example}%Real data analysis}
In this section, we utilized the proposed method to analyze a subset of data from 
the renowned Framingham heart study \citep{tsao2015} discussed in Section 1. 
The data set is available in the R package \emph{riskCommunicator}. Participants in this study have undergone biennial examinations since the study entry, and all subjects are continually monitored for cardiovascular outcomes. Our specific focus was on middle-aged patients aged between 30 and 50 years who were part of the first examination cycle.
We excluded subjects with a history of prevalent coronary heart disease, prevalent hypertension, myocardial infarction, or fatal coronary heart disease prior to the first examination. Additionally, subjects who passed away without experiencing any of these diseases were removed to avoid issues related to semi-competing risks. Missing observations were also excluded, resulting in a remaining sample size of 1753 patients in our analysis.

Researchers aimed to identify the effects of covariates on the occurrence of angina pectoris, myocardial infarction, coronary insufficiency, or fatal coronary heart disease (ANYCHD) and as well as hypertensive (HYPERTEN) events. The latter were defined as instances where high blood pressure was treated during the first examination or during the second examination when either the systolic blood pressure reached 140 mmHg or the diastolic blood pressure reached 90 mmHg. The survival times of interest were the time until the first ANYCHD event and the time until the first HYPERTEN event. The two times were measured in days and recorded from the same individual might be correlated. The bivariate times can either be observed directly or subjected to censoring due to death or loss of follow-up, resulting in a censoring rate of $62.3\%$.
The risk factors of interest included body mass index (BMI), systolic blood pressure (SYSBP, measured in mmHg), current cigarette smoking at the time of examination (CURSMOKE, yes= 1 and no= 0), sex (female= 1 and male= 0), and serum total cholesterol level (measured in mg/dL) in logarithmic transformation. Preliminary analysis indicated that these risk factors had no significant effects on censoring variables. Given that only $37.7\%$ of the survival times are observable, it's important to note that coefficients at quantile levels exceeding 0.4 cannot be reliably estimated. 

Supplementary Figures S.1-S.2 illustrate the comprehensive trajectories of coefficient estimations as $\tau$ increases with some particular values of $t_0$. In these figures, the black curves represent coefficient estimates, accompanied by their 95$\%$ RBS (red dashed curves) confidence intervals. At lower quantile levels and smaller $t_0$, RBS and CFS show similar trends. However, CFS estimator becomes unstable and unestimable for higher quantile levels and larger $t_0$, thus CFS estimator is omitted in Supplementary Figures S.1-S.2.
Table \ref{table:real} summarizes estimates of regression coefficients and their significance  as well as $\tau$-th conditional quantile of the logarithm of residual lifetime $\theta_{\tau,t_0}^{(k)}$ for selected %some specific 
patient $k$ for $k=1,2$ under $\tau=0.1,0.2,0.3$ quantile level and $t_0=0, 1200, 2400$ (days). Patient 1 is a female and non-smoker and has the minimum BMI, SYSBP, TOTCHOL among the sample, while Patient 2 is a female and smoker who has the maxmimum values of BMI, SYSBP and TOTCHOL.  
The IFR/RBS variance estimator with the number of replicates $B=500$ are used to compute the significance. It is noteworthy that the intercept exhibits a significant impact on event times. Moreover, both BMI and systolic blood pressure demonstrate significance, particularly at lower quantiles or for smaller values of $t_0$.  

\begin{figure}[h]
    \centering
    \includegraphics{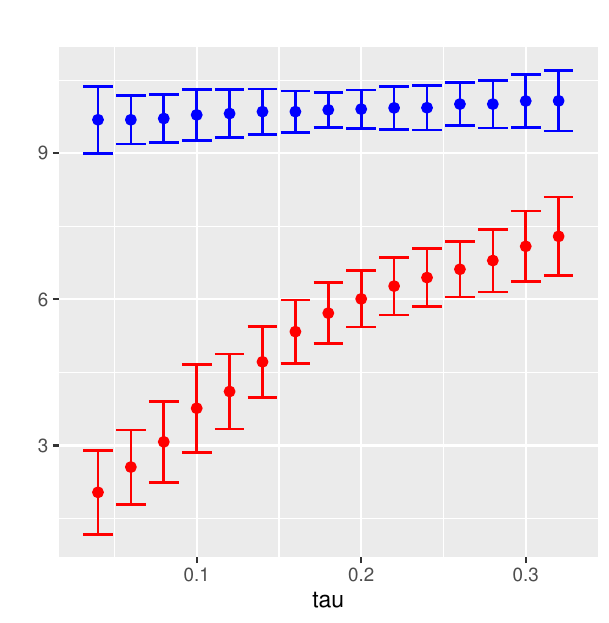}
    \caption{Prediction intervals of the logarithm residual lifetime with $t_0=1200$ over different quantile levels of $\tau$ for the selected patient 1 (in blue color) and patient 2 (in red color). % two specific patients.
    }
    \label{fig:realdata_prediction}
\end{figure}

Note that the estimates of $\theta_{\tau,t_0}^{(1)}$ in Table \ref{table:real} do not increase as $\tau$ increases, suggesting a crossing quantile problem in the analysis. 
Thus we further use a rearrangement procedure proposed by \cite{chernozhukov2010cross} to construct a monotone quantile curve, denoted by $\theta_{\tau,t_0}^{(k)*}$ in the table.
It can be seen from this table that patients with smoke hobby and higher values of BMI, SYSBP and TOTCHOL face higher risks and have shorter remaining time until the occurrence of severe cardiovascular diseases. 
Moreover, to illustrate the effects of the rearrangement procedure in prediction, we consider $t_0=1200$ and calculate the complete estimated $\theta_{\tau,t_0}^{(k)*}$ 
at different quantile levels $\tau$ for both selected 
patients. Figure \ref{fig:realdata_prediction} visualizes 
the prediction intervals at different quantile levels for the first and second patients, respectively. 
Notably, the difference between the two typical patients is quite large at small quantile levels, and lessens as quantile level increases.

\section{Discussion}

This article introduces a marginal QRL regression approach to  accommodate the potentially clustered  failure times
when there are multiple failure event types or groups of subjects in the study. The estimation process is computationally simple and stable, making it attractive for practical applications. Our proposed variance estimators in Section 3.2 are particularly tailored for the estimator $\widehat{\alpha}_N$, which is obtained by solving the estimation equation (6). These asymptotic variance estimators address the within-cluster dependence, making subsequent inference more reliable. This marginal approach is valuable when the relationship between quantile residual lifetimes and covariates is of interest, given that a subject is known to be disease-free at a specific time point. Our proposal leaves the underlying correlation structure completely unspecified, making it robust to potential misspecification and flexible in modeling various multivariate failure times. 

The estimating equation (6) is analogous to the well-known generalized estimating equations (GEE) approach with an independent working correlation structure. The GEE method has been extended to quantile regression for longitudinal data in the literature, such as \cite{jung1996quasi}, \cite{fu2012quantile} and \cite{leng2014smoothing}. We adopt the independent working model in light of the considerations as follows. 1) The choice of the working correlation structure should be a trade-off between simplicity and potential efficiency loss from misspecification. 2) Since the association is considered as nuisance in the marginal models, a simpler working correlation will generally suffice, with the independent working structure being recommended by \cite{fahrmeir2013multivariate}. Our simulation results demonstrate the promising performance of the proposed method across various dependence structures and copula types.

While we acknowledge that incorporating within-cluster dependence may improve efficiency, integrating the idea of the GEE approach within the framework of the multivariate quantile residual lifetime model poses challenges. 
As a potential direction for future work, we consider the following weighted estimating equations for residual lifetimes:
\begin{equation} \label{eq:R1}
    \frac{1}{n}\sum\limits_{i=1}^n \mathbf{X}_i^T\bm{\mathcal{W}}_i^{-1}\bm{\zeta}_i=\mathbf{0},
\end{equation}
where $\mathbf{X}_{i}=(\mathbf{X}_{i1},\cdots,\mathbf{X}_{im_i})^T$, $\bm{\zeta_i} = (\zeta_{i1},\cdots,\zeta_{im_i})^T$ with $$\zeta_{ij} = I(Y_{ij}\geq t_0)\left[\Delta_{ij}I\left\{Y_{ij}\leq t_0+\exp(\bfX_{ij}^T\bfalpha) \right\}\hatG(t_0)/\hatG(Y_{ij}) - \tau \right].$$
$\bm{\mathcal{W}}_i$ is a working covariance matrix of $\bm{\zeta}_i$ and can be expressed as $\bm{\mathcal{W}}_i=\bm{\Gamma}_i^{\frac{1}{2}}\mathbf{A}_i\bm{\Gamma}_i^{\frac{1}{2}}$, where $\bm{\Gamma}_i=\hbox{diag}\{\sigma^2_{i1},\cdots,\sigma^2_{im_i}\}$ with $\sigma^2_{ij}$ being the dispersion of $\zeta_{ij}$. $\mathbf{A}_i$ is a correlation matrix that can be specified with some unknown parameters or as a linear combination of some known basis matrices \citep{qu2000qif}. It is noted that potential issues may arise from \eqref{eq:R1} demanding more in-depth exploration. First, the dependence may vary with quantile levels or the time points $t_0$, making it difficult to specify a proper working correlation structure. Second, as $t_0$ increases, the number of individuals with $T_{ij}\geq t_0$ will decrease, and the unstable estimation may become more severe for larger $t_0$ if an inappropriate correlation structure is imposed.  Besides, the potential efficiency gains from incorporating a weight function require further investigation through 
theoretical justification and numerical studies.

Additionally, we assume the censoring variable $C_{ij}$'s are i.i.d from a distribution  independent from $\bfX_{ij}$. 
In practice, it may be necessary to verify this assumption about the censoring distribution before applying the proposed method.
Our method can be simply improved by incorporating covariates in modeling the censoring times through Cox proportional hazards model for example,  and replace 
$\hatG(\cdot)$ in \eqref{eq:estimating equation} with $\hatG(\cdot|X)$. Further study of its theoretical justification is also warranted. 

\section*{Supplementary Material}
The online Supplementary Material contains an appendix for technical proofs of the lemma and theorems referenced in Section 3 and additional numerical results referenced in Sections 4-5.

\section*{Acknowledgements}
We are grateful to the editor, the associate editor, and three referees for their valuable comments and suggestions, which have greatly improved the article.
This research was supported by the Singapore Ministry of Education Academic Research Fund Tier 2 Grant (MOE-T2EP20121-0004).

\bibliography{ref_qres}
\bibliographystyle{apalike}

\begin{table}[htbp]
\footnotesize
\caption{Estimation results based on 500 replicates for quantile level $\tau=0.5$ under Scenario 1 with Kendall's tau=0.5.} 
\label{table: sim1-2}
\begin{center}
\begin{tabular}{llllllrlllc}
\hline\hline
\multicolumn{3}{c}{}&\multicolumn{3}{c}{$\alpha_0(0.5,t_0)$}&\multicolumn{1}{c}{}&\multicolumn{3}{c}{$\alpha_1(0.5,t_0)$}&\multicolumn{1}{c}{runtime}\tabularnewline
\cline{4-6}\cline{8-10}
$(n,m)$&&&$t_0=0$&$t_0=1$&$t_0=2$&&$t_0=0$&$t_0=1$&$t_0=2$&(s)\tabularnewline
\hline
(200,3)&bias&&-0.007&-0.009&-0.005&&0.008&0.019&0.017\tabularnewline
&MCSD&&0.146&0.155&0.174&&0.235&0.259&0.289\tabularnewline
&ASE&IFR&0.130&0.150&0.173&&0.235&0.267&0.303\tabularnewline
&&FR&0.146&0.158&0.178&&0.24&0.268&0.303&5.75\tabularnewline
&&CFS&0.142&0.164&0.196&&0.234&0.276&0.332&0.262\tabularnewline
&&RBS&0.139&0.149&0.167&&0.227&0.249&0.281& 3.502\tabularnewline
&CP&IFR&0.924&0.932&0.942&&0.956&0.938&0.958\tabularnewline
&&FR&0.958&0.95&0.958&&0.96&0.958&0.964\tabularnewline
&&CFS&0.948&0.958&0.964&&0.956&0.948&0.98\tabularnewline
&&RBS&0.95&0.932&0.94&&0.946&0.944&0.94\tabularnewline
\cline{1-11}
(500,3)&bias&&0.002&0.003&0.001&&0.004&0.005&0.008\tabularnewline
&MCSD&&0.087&0.096&0.109&&0.138&0.161&0.186\tabularnewline
&ASE&IFR&0.082&0.094&0.108&&0.147&0.166&0.188\tabularnewline
&&FR&0.09&0.099&0.11&&0.148&0.167&0.188&9.948\tabularnewline
&&CFS&0.09&0.104&0.122&&0.146&0.173&0.206&0.673\tabularnewline
&&RBS&0.087&0.096&0.106&&0.142&0.161&0.178&6.265\tabularnewline
&CP&IFR&0.920&0.936&0.952&&0.940&0.950&0.963\tabularnewline
&&FR&0.956&0.956&0.948&&0.958&0.96&0.942\tabularnewline
&&CFS&0.95&0.954&0.978&&0.938&0.962&0.98\tabularnewline
&&RBS&0.946&0.95&0.938&&0.954&0.95&0.938\tabularnewline
\cline{1-11}
(200,10)&bias&&-0.007&-0.003&0&&0.004&0.001&-0.002\tabularnewline
&MCSD&&0.104&0.099&0.105&&0.129&0.144&0.163\tabularnewline
&ASE&IFR&0.070&0.081&0.092&&0.127&0.143&0.159\tabularnewline
&&FR&0.1&0.102&0.106&&0.129&0.147&0.165&10.978\tabularnewline
&&CFS&0.099&0.105&0.116&&0.126&0.151&0.181&0.825\tabularnewline
&&RBS&0.096&0.098&0.103&&0.125&0.141&0.159&6.363\tabularnewline
&CP&IFR&0.868&0.906&0.946&&0.946&0.952&0.960\tabularnewline
&&FR&0.954&0.962&0.95&&0.954&0.95&0.944\tabularnewline
&&CFS&0.952&0.972&0.97&&0.946&0.966&0.972\tabularnewline
&&RBS&0.948&0.948&0.944&&0.948&0.942&0.932\tabularnewline
\hline
\end{tabular}\end{center}
\end{table}

\begin{table}[htbp]
\footnotesize
\caption{Estimation results based on 500 replicates for quantile level $\tau=0.5$ under Scenario 2.}
\label{table: sim2}
\begin{center}
\begin{tabular}{llllllrlllc}
\hline\hline
\multicolumn{3}{c}{}&\multicolumn{3}{c}{$\alpha_0(0.5,t_0)$}&\multicolumn{1}{c}{}&\multicolumn{3}{c}{$\alpha_1(0.5,t_0)$}&\multicolumn{1}{c}{runtime}\tabularnewline
\cline{4-6}\cline{8-10} $(n,m)$&&&$t_0=0$&$t_0=1$&$t_0=2$&&$t_0=0$&$t_0=1$&$t_0=2$&(s)\tabularnewline
\hline
(200,3)&bias&&-0.008&-0.002&-0.014&&0.003&0.001&0.006&\tabularnewline
&MCSD&&0.174&0.19&0.191&&0.313&0.33&0.334&\tabularnewline
&ASE&IFR&0.131&0.15&0.173&&0.238&0.269&0.303\tabularnewline
&&FR&0.179&0.184&0.164&&0.317&0.328&0.291&9.672\tabularnewline
&&CFS&0.178&0.191&0.179&&0.314&0.34&0.315&0.358\tabularnewline
&&RBS&0.169&0.173&0.154&&0.295&0.306&0.268&5.806\tabularnewline
&CP&IFR&0.848&0.90&0.929&&0.844&0.894&0.927\tabularnewline
&&FR&0.948&0.94&0.908&&0.948&0.944&0.925&\tabularnewline
&&CFS&0.948&0.948&0.939&&0.946&0.95&0.946&\tabularnewline
&&RBS&0.936&0.926&0.892&&0.936&0.93&0.894&\tabularnewline
\hline
(500,3)&bias&&-0.009&-0.009&-0.011&&0.011&0.012&0.013&\tabularnewline
&MCSD&&0.109&0.111&0.123&&0.197&0.195&0.212&\tabularnewline
&ASE&IFR&0.082&0.093&0.108&&0.148&0.166&0.188\tabularnewline
&&FR&0.112&0.116&0.118&&0.198&0.205&0.208&13.584\tabularnewline
&&CFS&0.111&0.121&0.129&&0.196&0.213&0.227&0.905\tabularnewline
&&RBS&0.107&0.111&0.114&&0.188&0.195&0.199&7.001\tabularnewline
&CP&IFR&0.868&0.896&0.909&&0.850&0.908&0.909\tabularnewline
&&FR&0.946&0.966&0.942&&0.954&0.96&0.952&\tabularnewline
&&CFS&0.946&0.972&0.963&&0.952&0.964&0.969&\tabularnewline
&&RBS&0.932&0.956&0.927&&0.942&0.952&0.944&\tabularnewline
\hline
(200,10)&bias&&-0.003&-0.002&-0.006&&0.002&-0.003&0.002&\tabularnewline
&MCSD&&0.166&0.148&0.142&&0.283&0.264&0.256&\tabularnewline
&ASE&IFR&0.071&0.081&0.094&&0.127&0.143&0.163\tabularnewline
&&FR&0.156&0.147&0.132&&0.274&0.262&0.239&14.968\tabularnewline
&&CFS&0.155&0.153&0.145&&0.272&0.273&0.262&0.879\tabularnewline
&&RBS&0.151&0.142&0.128&&0.263&0.254&0.23&7.404\tabularnewline
&CP&IFR&0.606&0.692&0.780&&0.612&0.708&0.766\tabularnewline
&&FR&0.926&0.938&0.927&&0.938&0.954&0.936&\tabularnewline
&&CFS&0.926&0.946&0.951&&0.934&0.964&0.953\tabularnewline
&&RBS&0.92&0.922&0.91&&0.92&0.94&0.925&\tabularnewline
\hline
\end{tabular}\end{center}
\end{table}

\begin{table}[htbp]
\footnotesize
\caption{Estimation results based on 500 replicates for quantile level $\tau=0.5$ under %the third scenario.
Scenario 3.}
\label{table: sim3}
\begin{center}
\begin{tabular}{llllllrlllc}
\hline\hline
\multicolumn{3}{c}{}&\multicolumn{3}{c}{$\alpha_0(0.5,t_0)$}&\multicolumn{1}{c}{}&\multicolumn{3}{c}{$\alpha_1(0.5,t_0)$}&\multicolumn{1}{c}{runtime}\tabularnewline
\cline{4-6}\cline{8-10} $(n,m)$&&&$t_0=0$&$t_0=1$&$t_0=2$&&$t_0=0$&$t_0=1$&$t_0=2$&(s)\tabularnewline
\hline
(200,3)&bias&&0.003&0.004&-0.003&$$&-0.01&-0.015&-0.009&\tabularnewline
&MCSD&&0.088&0.119&0.165&$$&0.161&0.217&0.296&\tabularnewline
&ASE&IFR&0.069&0.094&0.133&$$&0.123&0.165&0.235&9.662\tabularnewline
&&FR&0.092&0.126&0.167&$$&0.163&0.221&0.295&10.724\tabularnewline
&&CFS&0.092&0.131&0.185&$$&0.161&0.229&0.324&0.445\tabularnewline
&&RBS&0.09&0.12&0.161&$$&0.158&0.212&0.283&6.031\tabularnewline
&CP&IFR&0.884&0.888&0.884&$$&0.856&0.848&0.874&\tabularnewline
&&FR&0.968&0.966&0.956&$$&0.952&0.948&0.936&\tabularnewline
&&CFS&0.968&0.972&0.974&$$&0.948&0.962&0.96&\tabularnewline
&&RBS&0.958&0.962&0.95&$$&0.94&0.928&0.93&\tabularnewline
\hline
(500,3)&bias&&-0.001&-0.003&-0.004&$$&0.003&0.005&0.006&\tabularnewline
&MCSD&&0.057&0.077&0.1&$$&0.1&0.136&0.175&\tabularnewline
&ASE&IFR&0.043&0.057&0.082&$$&0.075&0.101&0.142&21.31\tabularnewline
&&FR&0.057&0.077&0.104&$$&0.1&0.135&0.18&23.796\tabularnewline
&&CFS&0.057&0.082&0.115&$$&0.1&0.143&0.2&1.286\tabularnewline
&&RBS&0.056&0.076&0.101&$$&0.098&0.133&0.175&12.957\tabularnewline
&CP&IFR&0.852&0.856&0.88&$$&0.852&0.846&0.892&\tabularnewline
&&FR&0.946&0.946&0.958&$$&0.95&0.952&0.954&\tabularnewline
&&CFS&0.948&0.962&0.97&$$&0.95&0.968&0.97&\tabularnewline
&&RBS&0.94&0.94&0.952&$$&0.944&0.946&0.95&\tabularnewline
\hline
(200,10)&bias&&0&-0.001&-0.003&$$&-0.005&-0.008&-0.007&\tabularnewline
&MCSD&&0.075&0.102&0.131&$$&0.131&0.18&0.234&\tabularnewline
&ASE&IFR&0.037&0.05&0.071&$$&0.065&0.087&0.124&31.27\tabularnewline
&&FR&0.079&0.108&0.136&$$&0.138&0.189&0.237&35.872\tabularnewline
&&CFS&0.079&0.114&0.152&$$&0.138&0.198&0.264&1.609\tabularnewline
&&RBS&0.077&0.105&0.132&$$&0.135&0.183&0.231&13.949\tabularnewline
&CP&IFR&0.696&0.688&0.734&$$&0.706&0.684&0.722&\tabularnewline
&&FR&0.966&0.964&0.95&$$&0.948&0.948&0.944&\tabularnewline
&&CFS&0.966&0.972&0.974&$$&0.948&0.96&0.966&\tabularnewline
&&RBS&0.962&0.958&0.942&$$&0.936&0.932&0.944&\tabularnewline
\hline
\end{tabular}\end{center}
\end{table}

\begin{table}[htbp]
\footnotesize
\tabcolsep=3pt
\caption{Estimation results based on 500 replicates under Scenario 4 ($n=200,m = 10$).} 
\label{table: sim4}
%simdata=6
\begin{center}
\begin{tabular}{llllllrlllrlllrlll}
\hline\hline
\multicolumn{3}{c}{}&\multicolumn{3}{c}{$\alpha_0(0.25,t_0)$}&\multicolumn{1}{c}{}&\multicolumn{3}{c}{$\alpha_1(0.25,t_0)$}&&\multicolumn{3}{c}{$\alpha_0(0.5,t_0)$}&\multicolumn{1}{c}{}&\multicolumn{3}{c}{$\alpha_1(0.5,t_0)$}\tabularnewline
\cline{4-6}\cline{8-10}\cline{12-14}\cline{16-18}
$a$&&&$t_0=0$&$t_0=1$&$t_0=2$&&$t_0=0$&$t_0=1$&$t_0=2$&&$t_0=0$&$t_0=1$&$t_0=2$&&$t_0=0$&$t_0=1$&$t_0=2$\tabularnewline
\hline
0.1& truth&&-0.939&-0.939&-0.939&&2.194&2.127&2.09 &&-0.06&-0.06&-0.06&&2.106&2.068&2.044\tabularnewline
&bias&&-0.009&-0.005&-0.003&&-0.003&-0.007&-0.01&&-0.005&-0.008&0.002&&-0.007&-0.003&-0.013\tabularnewline
&MCSD&&0.16&0.118&0.15&&0.211&0.188&0.218&&0.102&0.09&0.108&&0.142&0.14&0.157\tabularnewline
&ASE&IFR&0.064&0.095&0.144&&0.089&0.118&0.163&&0.047&0.07&0.106&&0.068&0.089&0.122\tabularnewline
&&FR&0.166&0.116&0.147&&0.226&0.186&0.203&&0.109&0.091&0.109&&0.15&0.138&0.15\tabularnewline
&&CFS&0.166&0.123&0.162&&0.226&0.198&0.225&&0.109&0.097&0.12&&0.15&0.146&0.166\tabularnewline
&&RBS&0.155&0.111&0.138&&0.213&0.178&0.191&&0.106&0.088&0.105&&0.145&0.133&0.145\tabularnewline
&CP&IFR&0.552&0.89&0.936&&0.59&0.768&0.848&&0.642&0.878&0.932&&0.672&0.782&0.872\tabularnewline
&&FR&0.954&0.944&0.946&&0.964&0.952&0.94&&0.97&0.95&0.94&&0.962&0.956&0.94\tabularnewline
&&CFS&0.954&0.952&0.966&&0.964&0.96&0.966&&0.97&0.96&0.966&&0.962&0.96&0.964\tabularnewline
&&RBS&0.938&0.936&0.928&&0.952&0.946&0.92&&0.962&0.94&0.93&&0.956&0.95&0.928\tabularnewline
\hline
0.2&truth&&-0.939&-0.939&-0.939&&2.388&2.276&2.202&&-0.06&-0.06&-0.06&&2.212&2.148&2.101\tabularnewline
&bias&&-0.009&-0.005&-0.003&&-0.003&-0.007&-0.01&&-0.005&-0.008&0.002&&-0.005&-0.002&-0.012\tabularnewline
&MCSD&&0.16&0.118&0.15&&0.201&0.183&0.213&&0.102&0.09&0.108&&0.135&0.134&0.153\tabularnewline
&ASE&IFR&0.064&0.095&0.144&&0.086&0.115&0.161&&0.047&0.07&0.106&&0.065&0.086&0.12\tabularnewline
&&FR&0.165&0.116&0.147&&0.215&0.181&0.201&&0.109&0.091&0.109&&0.143&0.132&0.147\tabularnewline
&&CFS&0.166&0.123&0.162&&0.216&0.192&0.223&&0.109&0.097&0.12&&0.143&0.14&0.162\tabularnewline
&&RBS&0.156&0.111&0.138&&0.204&0.174&0.191&&0.106&0.087&0.105&&0.138&0.127&0.141\tabularnewline
&CP&IFR&0.554&0.89&0.938&&0.596&0.766&0.864&&0.642&0.876&0.932&&0.672&0.802&0.866\tabularnewline
&&FR&0.954&0.942&0.946&&0.97&0.944&0.94&&0.972&0.948&0.94&&0.956&0.954&0.948\tabularnewline
&&CFS&0.954&0.952&0.966&&0.97&0.95&0.96&&0.972&0.962&0.966&&0.956&0.96&0.97\tabularnewline
&&RBS&0.938&0.936&0.928&&0.956&0.936&0.932&&0.962&0.938&0.928&&0.952&0.944&0.94\tabularnewline
\hline
0.5&truth&&-0.939&-0.939&-0.939&&2.97&2.839&2.71&&-0.06&-0.06&-0.06&&2.53&2.445&2.361\tabularnewline
&bias&&-0.009&-0.005&-0.003&&0.001&-0.004&-0.008&&-0.005&-0.008&0.002&&-0.002&0.001&-0.01\tabularnewline
&MCSD&&0.16&0.118&0.15&&0.176&0.152&0.185&&0.102&0.09&0.107&&0.117&0.111&0.131\tabularnewline
&ASE&IFR&0.064&0.095&0.144&&0.075&0.105&0.153&&0.047&0.07&0.106&&0.057&0.079&0.113\tabularnewline
&&FR&0.165&0.116&0.147&&0.188&0.152&0.183&&0.109&0.091&0.109&&0.125&0.112&0.13\tabularnewline
&&CFS&0.166&0.123&0.162&&0.188&0.161&0.203&&0.109&0.097&0.12&&0.125&0.119&0.143\tabularnewline
&&RBS&0.156&0.111&0.139&&0.177&0.146&0.174&&0.105&0.088&0.105&&0.12&0.108&0.126\tabularnewline
&CP&IFR&0.554&0.89&0.936&&0.584&0.838&0.894&&0.642&0.876&0.932&&0.686&0.858&0.912\tabularnewline
&&FR&0.954&0.942&0.944&&0.968&0.94&0.956&&0.972&0.948&0.942&&0.966&0.952&0.954\tabularnewline
&&CFS&0.954&0.952&0.966&&0.968&0.954&0.978&&0.972&0.962&0.968&&0.966&0.958&0.974\tabularnewline
&&RBS&0.938&0.936&0.928&&0.958&0.926&0.94&&0.962&0.938&0.93&&0.954&0.938&0.942\tabularnewline
\hline
\end{tabular}\end{center}
\end{table}

\begin{table}[htbp]
\begin{center}
\caption{Estimation of regression coefficients and quantile of the residual lifetime $\theta_{\tau,t_0}^{(k)}$ ($k=1,2$) for the Framingham heart data
with $\tau=0.1,0.2,0.3$ quantiles of  ANYCHD/HYPERTEN times after the first examination at $t_0=0, 1200, 2400$ (days), respectively. }
\label{table:real}
\scriptsize
\begin{tabular}{lllllllllllll}
\hline\hline
\multicolumn{1}{l}{}&\multicolumn{1}{l}{}&\multicolumn{3}{c}{$t_0=0$}&\multicolumn{3}{c}{$t_0=1200$}&&\multicolumn{3}{c}{$t_0=2400$}\tabularnewline
\cline{3-5}\cline{7-9}\cline{11-13}
&&$\tau=0.1$&$\tau=0.2$&$\tau=0.3$&&$\tau=0.1$&$\tau=0.2$&$\tau=0.3$&&$\tau=0.1$&$\tau=0.2$&$\tau=0.3$\tabularnewline
\hline
Estimates\tabularnewline
Intercept&&17.046&14.916&13.862&&17.808&14.218&13.071&&12.641&12.842&11.446\tabularnewline
BMI&&-0.045&-0.039&-0.023&&-0.076&-0.034&-0.023&&-0.021&-0.016&-0.011\tabularnewline
SYSBP&&-0.045&-0.038&-0.029&&-0.046&-0.034&-0.022&&-0.033&-0.024&-0.014\tabularnewline
CURSMOKE&&0.061&0.01&-0.026&&0.082&-0.026&-0.051&&-0.113&-0.086&-0.053\tabularnewline
SEX&&0.07&0.077&0.015&&-0.014&0.131&0.035&&0.253&0.071&0.049\tabularnewline
log(TOTCHOL)&&-0.512&-0.196&-0.197&&-0.536&-0.202&-0.217&&-0.093&-0.215&-0.158\tabularnewline
$\theta_{\tau,t_0}^{(1)}$&&10.125&10.171&10.075&&10.07&9.932&9.813&&9.295&9.543&9.422\tabularnewline
$\theta_{\tau,t_0}^{(1)*}$&&9.973&10.105&10.163&&9.785&9.904&10.07&&9.295&9.473&9.543\tabularnewline
$\theta_{\tau,t_0}^{(2)}$&&4.699&5.836&6.824&&3.763&6.008&7.087&&5.846&6.762&7.786\tabularnewline
$\theta_{\tau,t_0}^{(2)*}$&&4.699&5.836&6.824&&3.763&6.008&7.087&&5.846&6.762&7.786\tabularnewline
\hline
SE- RBS\tabularnewline
(Intercept)&&1.161**&0.982**&1.101**&&1.686**&1.093**&1.391**&&1.382**&1.357**&1.934**\tabularnewline
BMI&&0.01**&0.009**&0.008**&&0.017**&0.011**&0.012*&&0.013&0.013&0.016\tabularnewline
SYSBP&&0.003**&0.003**&0.004**&&0.004**&0.004**&0.005**&&0.004**&0.004**&0.009\tabularnewline
CURSMOKE&&0.078&0.066&0.061&&0.104&0.068&0.068&&0.092&0.069&0.112\tabularnewline
SEX&&0.069&0.066&0.056&&0.106&0.069*&0.075&&0.091**&0.071&0.106\tabularnewline
log(TOTCHOL)&&0.226**&0.185&0.172&&0.292*&0.188&0.208&&0.266&0.215&0.285\tabularnewline
$\theta_{\tau,t_0}^{(1)}$&&0.158**&0.151**&0.218**&&0.267**&0.2**&0.276**&&0.216**&0.234**&0.439**\tabularnewline
$\theta_{\tau,t_0}^{(2)}$&&0.231**&0.204**&0.267**&&0.461**&0.295**&0.369**&&0.317**&0.339**&0.415**\tabularnewline
\hline
SE- IFR\tabularnewline
(Intercept)&&1.118**&0.742**&0.562**&&1.387**&0.761**&0.617**&&1.47**&0.907**&0.539**\tabularnewline
BMI&&0.01**&0.008**&0.006**&&0.016**&0.011**&0.009**&&0.014&0.011&0.006*\tabularnewline
SYSBP&&0.003**&0.002**&0.002**&&0.004**&0.002**&0.002**&&0.004**&0.002**&0.002**\tabularnewline
CURSMOKE&&0.077&0.053&0.033&&0.109&0.053&0.039&&0.114&0.052*&0.028*\tabularnewline
SEX&&0.079&0.062&0.04&&0.115&0.068*&0.053&&0.095**&0.072&0.07\tabularnewline
log(TOTCHOL)&&0.221**&0.143&0.105*&&0.282*&0.146&0.117*&&0.288&0.166&0.085*\tabularnewline
$\theta_{\tau,t_0}^{(1)}$&&0.134**&0.101**&0.076**&&0.182**&0.104**&0.081**&&0.207**&0.114**&0.067\tabularnewline
$\theta_{\tau,t_0}^{(2)}$&&0.231**&0.187**&0.138**&&0.34**&0.229**&0.175**&&0.324**&0.218**&0.185**\tabularnewline
\hline
\multicolumn{12}{l}{* and ** indicate significance at levels 0.1 and 0.05, respectively. The significance is computed based on}\tabularnewline
\multicolumn{12}{l}{RBS/IFR variance estimators.}\tabularnewline
\end{tabular}
\end{center}
\end{table}

\clearpage
\appendix
\renewcommand{\theequation}{A.\arabic{equation}} % This line ads "A."
\renewcommand{\thetable}{S.\arabic{table}}
\renewcommand{\thefigure}{S.\arabic{figure}}
\setcounter{table}{0}
\setcounter{figure}{0}
\setcounter{equation}{0}
\begin{center}
\Large{\textbf{Supplementary Material for \\
``Quantile Residual Lifetime Regression for Multivariate Failure Time Data"}}
\end{center}
\section{Technical Proofs}
\begin{proof}[Proof of Theorem 1]
The strong consistency of $\hatalpha_N$ can obtained from  L\'{e}vy's theorem \citep{resnick2019prob}, Lemma 2.2 of \cite{white1980nonlinear} and Theorem 5.9 of \cite{van2000asymptotic}. In detail, it remains to verify
1) $||S_N(\bfalpha) - \barS_N(\bfalpha)|| \rightarrow_{a.s.} 0 $ uniformly for all $\bfalpha\in \mathcal{D}$;
2) $\inf\limits_{\bfalpha: ||\bfalpha - \bfalpha_0|| = \epsilon} ||\barS_N(\bfalpha)||$ is strictly positive for any $\epsilon>0$. By Conditions 1-2, the proof of the former result is similar to the proof of consistency under an independent case, readers can refer to the appendix in \cite{li2016quantile}.
We now show the latter result.
From the model (3) and the fact that $\Pr\left\{ T_{ij}\geq t_0 + \exp(\bfX_{ij}^T\bfalpha_0)|\bfX_{ij}\right\} - (1-\tau) \Pr\left\{T_{ij}\geq t_0|\bfX_{ij}\right\} = 0$, we can see that
\begin{equation*}
\begin{split}
&\Pr\left\{ T_{ij}\geq t_0 + \exp(\bfX_{ij}^T\bfalpha)|\bfX_{ij}\right\} - (1-\tau) \Pr\left\{T_{ij}\geq t_0|\bfX_{ij}\right\}\\
& = \Pr\left\{ T_{ij}\geq t_0 + \exp(\bfX_{ij}^T\bfalpha)|\bfX_{ij}\right\} - \Pr\left\{ T_{ij}\geq t_0 + \exp(\bfX_{ij}^T\bfalpha_0)|\bfX_{ij}\right\}\\
& = \Pr( e_{ij}^{\tau}\geq \bfX_{ij}^T(\bfalpha - \bfalpha_0)|\bfX_{ij})-
\Pr( e_{ij}^{\tau}\geq 0|\bfX_{ij})\\
& = F_e( 0|\bfX_{ij})- F_e( \bfX_{ij}^T(\bfalpha - \bfalpha_0)|\bfX_{ij})\\
& = f_e( 0|\bfX_{ij})\bfX_{ij}^T(\bfalpha - \bfalpha_0).
\end{split}
\end{equation*}
Thus, by Conditions 1 and 3, 
$\inf\limits_{\bfalpha: ||\bfalpha - \bfalpha_0|| = \epsilon} ||\barS_N(\bfalpha)||> 0 $ for any $\epsilon>0$. We now have completed the proof of the consistency of the proposed estimator under the multivariate setting.

\end{proof}

\begin{proof}[Proof of Lemma 1]
$N^{1/2}S_N(\bfalpha_0)$ can be expressed as
\begin{equation}
\begin{split}
& N^{1/2}S_N(\bfalpha_0)
= N^{-1/2}\sum\limits_i\psi_i(\bfalpha_0)\\
&+N^{-1/2}\sum\limits_i\sum\limits_j \bfX_{ij}\Delta_{ij}I\left\{t_0\leq Y_{ij}\leq t_0+\exp(\bfX_{ij}^T\bfalpha_0) \right\}\Bigg[\frac{\hatG(t_0)}{\hatG(Y_{ij})}-
\frac{G(t_0)}{G(Y_{ij})}\Bigg],
\end{split}
\label{eqpf:S_n}
\end{equation}
where
\begin{equation*}
\psi_i(\bfalpha)= \sum\limits_{j=1}^{m_i} \bfX_{ij}I(Y_{ij}\geq t_0)\left[\frac{\Delta_{ij}I\left\{Y_{ij}\leq t_0+\exp(\bfX_{ij}^T\bfalpha) \right\}}{G(Y_{ij})/G(t_0)} - \tau \right].
\end{equation*}
The second term in \eqref{eqpf:S_n} involves the estimator of $G$ and can be represented by virtue of martingale processes. We see from \citep{fleming2013counting} that
\begin{align}	
\widehat{G}_{n}(s)-G(s)= -G(s)\dfrac{1}{N}\sum_{i}\int_{-\infty}^{s}\dfrac{dM_{i}(u)}{y(u)}+o(1)
\label{eq:estG_aymptotic_represent}
\end{align}	
where $M_{i}(u)=\sum_{j}I(Y_{ij}\leq u,\Delta_{ij} = 0) - \int_{0}^{u}I(Y_{ij}\geq t)d\wedge_G(t)$ and at-risk indicator $y(u)=\lim_{n \rightarrow \infty} N^{-1}\sum_{i,j}I(Y_{ij}\ge u)$, $\bm\wedge_{G}(u)$ is the cumulative hazard function of the censoring variables $C_{ij}$.
Through simple algebraic manipulations, \eqref{eqpf:S_n} can be written as
\begin{equation}
N^{1/2}S_N(\bfalpha_0)=N^{-1/2}\sum\limits_i\bfomega_i(\bfalpha_0)+o(1),
\end{equation}
where $\bfomega_i(\bfalpha) = \psi_i(\bfalpha)+ \eta_i(\bfalpha)$, and
\begin{equation*}
\begin{split}
\eta_i(\bfalpha_0)
& = \sum\limits_{k,j} \frac{\bfX_{kj}\delta_{kj}I\left\{t_0\leq Y_{kj}\leq t_0+\exp(\bfX_{kj}^T\bfalpha_0) \right\}}{N G(Y_{kj})/G(t_0)}\int_{t_0}^{Y_{kj}}\dfrac{dM_{i}(u)}{y(u)}.
\end{split}
\end{equation*}

By Condition 1,  the Lyapunov central limit theorem and martingale central limit theorem, the  distribution of $N^{1/2}S_N(\bfalpha_0)$ is asymptotically normal with mean
zero and covariance matrix $\Sigma = N^{-1}\sum_{i=1}^n \bfomega_i(\bfalpha_0)\bfomega_i(\bfalpha_0)^T$.

\end{proof}

\begin{proof}[Proof of Theorem 2]
As seen from the proof of Lemma 1, $NS_N(\bfalpha)$ is asymptotically equivalent to sums of independent random vectors, $\bfomega_i(\bfalpha)$, we then explore the limiting distribution of $\hatalpha_N$ based on the asymptotic properties of $\bfomega_i(\bfalpha)=\psi_i(\bfalpha)+\eta_i(\bfalpha)$.
Since $\hatalpha_N$ satisfies $S_N(\hatalpha_N) = o(N^{-1/2})$, we see that
\begin{equation*}
\begin{split}
&\sum\limits_{i=1}^n \psi_i(\hatalpha_N)\\
&= \sum\limits_{i=1}^n\sum\limits_{j=1}^{m_i} \bfX_{ij}I(Y_{ij}\geq t_0)\left[\frac{\Delta_{ij}I\left\{Y_{ij}\leq t_0+\exp(\bfX_{ij}^T\hatalpha_N) \right\}}{G(Y_{ij})/G(t_0)} - \tau \right]\\
& = o(N^{1/2})+\sum\limits_i\sum\limits_j \bfX_{ij}\Delta_{ij}I\left\{t_0\leq Y_{ij}\leq t_0+\exp(\bfX_{ij}^T\hatalpha_N) \right\}\Bigg[\frac{G(t_0)}{G(Y_{ij})}-\frac{\hatG(t_0)}{\hatG(Y_{ij})}\Bigg]\\
& = o(N^{1/2})-\sum\limits_{i=1}^n \eta_i(\hatalpha_N)+o(1),
\end{split}
\end{equation*}
where the second equality is derived from  substituting $S_N(\hatalpha_N)$ into, the third is from  Equation \eqref{eqpf:S_n}.
Note that $\bfomega_i(\bfalpha)$'s are independent and $\sum\limits_{i=1} E\bfomega_i(\bfalpha) = N\barS_N(\bfalpha)$.
Thus, by Condition 
4, we have $\sum\limits_{i=1}^n \bfomega_i(\hatalpha_N) = o(n^{1/2})$,
which is same as (2.1) in \cite{he1996Bahadur}. Using the similar techniques in \cite{wang2009inference}, we also denote
$u_i(\bfalpha,d) = \sup\limits_{||\bfgamma - \bfalpha||\leq d}||\bfomega_i(\bfgamma) - \bfomega_i(\bfalpha)||$. According to  \cite{he1996Bahadur}, it remains to show conditions B3-B4, B5' and B8 in their Theorem 2 are satisfied under our setting.

B3. Denote $\tildee_{ij}(\bfalpha) = \log(Y_{ij}-t_0) - \bfX_{ij}^T\bfalpha$,$e_{ij}(\bfalpha) = \log(T_{ij}-t_0) - \bfX_{ij}^T\bfalpha$. By the fact that $\sup\limits_{t\leq t_u}|\hatG(t)-G(t)|=o(N^{-1/2+\epsilon})$ for any $\epsilon>0$, we see that for $||\bfalpha-\bfalpha_0||\leq d_0$ and $d\leq d_0$,
\begin{equation*}
\begin{split}
&u_i(\bfalpha,d)\\
&=\sup\limits_{||\bfgamma - \bfalpha||\leq d} \Bigg|\Bigg| \sum\limits_j \bfX_{ij}\Delta_{ij}\delta_{ij}I(T_{ij}\geq t_0)\frac{\hatG(t_0)}{\hatG(Y_{ij})}\Bigg[I\left\{e_{ij}(\alpha)\leq \bfX_{ij}^T(\bfgamma-\bfalpha)  \right\}-I\left\{e_{ij}(\alpha)\leq 0  \right\}\Bigg]\Bigg|\Bigg| \\
&\leq c_1\sum\limits_j ||\bfX_{ij}||\Delta_{ij}I(T_{ij}\geq t_0)\frac{G(t_0)}{G(Y_{ij})}\sup\limits_{||\bfgamma - \bfalpha||\leq d}I\left\{e_{ij}(\bfalpha)\leq \max(0,\bfX_{ij}^T(\bfgamma-\bfalpha) \right\}\\
&\leq c_1\sum\limits_j ||\bfX_{ij}||\Delta_{ij}I(T_{ij}\geq t_0)\frac{G(t_0)}{G(Y_{ij})}I\left\{e_{ij}(\bfalpha)\leq d||\bfX_{ij}|| \right\}\\
&:= c_1B_i,
\end{split}
\end{equation*}
for some constant $c_1$.
From Condition 1, there exists some constant $c_2$ such that
\begin{equation*}
\begin{split}
E(B_i^2|\bfX_{ij}, T_{ij}\geq t_0)& \leq  n_i\left(G(t_0)\right)^2 \sum\limits_{j} ||\bfX_{ij}||^2 \Pr\left\{e_{ij}(\alpha)\leq d||\bfX_{ij}|| \right\}/G(Y_{ij})\\
&\leq c_2 n_i\sum\limits_{j} ||\bfX_{ij}||^2\Pr\left\{e_{ij}(\alpha)\leq d||\bfX_{ij}|| \right\}\\
&\leq c_2 n_i\sum\limits_{j} ||\bfX_{ij}||^2\left[1+f_e(0|\bfX_{ij})d||\bfX_{ij}||\right]+ O(d^2)
\end{split}
\end{equation*}

Then, results in Condition B3 of \cite{he1996Bahadur} follows by fixing
$a_i=c_3 d^{-1}\sqrt{n_i\sum\limits_{j} ||\bfX_{ij}||^2 \left[1+f_e(0|\bfX_{ij})d||\bfX_{ij}||\right]}$ for some constant $c_3$.

B4. Under Conditions 3-4, $A_n = \sum\limits_i a_i = O(n)$, Condition B4 of \cite{he1996Bahadur}  can be obtained.
Condition B5' follows by taking $d_n = n^{-1/4}\log n$.

B8. Taking  the Taylor expansion of $\barS_N(\bfalpha)$ at $\bfalpha_0$ yields
\begin{equation*}
N^{1/2}(\barS_N(\hatalpha_N) - \barS_N(\bfalpha_0)) = N^{1/2}(\hatalpha_N - \bfalpha_0) \Lambda G(t_0) + o(1),
\end{equation*}
where $\Lambda$ is defined in Condition 3. Consequently, Condition B8 of \cite{he1996Bahadur} holds.

Therefore, all conditions in Theorem 2.2 of \cite{he1996Bahadur} hold, and we can apply their theorem, yielding the Bahadur representation in the form of
\begin{equation}
\begin{split}
 &N^{1/2}(\hatalpha_N-\bfalpha_0) = -N^{-1/2}\Lambda^{-1}(G(t_0))^{-1}\sum\limits_{i=1}\bfomega_i(\bfalpha_0)+o_p(1)
 \label{eq:alpha_aymptotic_represent}
\end{split}
\end{equation}
Thus, by the Lyapunov central limit theorem, $N^{1/2}(\hatalpha_N-\bfalpha)$ is asymptotically normal with covariance matrix as
$\operatorname{Cov}\left\{N^{1/2}(\hatalpha_N-\bfalpha)\right\}  = \widetilde{\Lambda}^{-1}\Sigma\widetilde{\Lambda}^{-1}$, with $\widetilde{\Lambda}= G(t_0)\Lambda$.

\end{proof}

\begin{proof}[Proof of Theorem 3]
It can be straightforwardly verified that the difference between the perturbed $G^{*}$ and the truth $G$ can be asymptotically represented by sums of independent random processes in analogy to Eq\eqref{eq:estG_aymptotic_represent}.
Then, given $E(\gamma_i) = 1$, following the arguments in proofs of Lemma 1 and Theorem 2, we can similarly get
\begin{equation}
N^{1/2}(\hatalpha^{*}-\bfalpha_0) = -N^{-1/2}\Lambda^{-1}(G(t_0))^{-1}\sum\limits_{i=1}\gamma_i\bfomega_i(\bfalpha_0)+o_p(1).
\end{equation}
When coupled with Eq \eqref{eq:alpha_aymptotic_represent}, this implies that
\begin{equation}
N^{1/2}(\hatalpha^{*}-\hatalpha_N) = -N^{-1/2}\Lambda^{-1}(G(t_0))^{-1}\sum\limits_{i=1}(\gamma_i-1)\bfomega_i(\bfalpha_0)+o_p(1).
\end{equation}
Since $\operatorname{Var}(\gamma_i)=1$, we have
\begin{equation*}
\begin{split}
&\operatorname{Cov}\left[N^{-\frac{1}{2}}\sum\limits_{i=1}(\gamma_i-1)\bfomega_i(\bfalpha_0)\Bigg|\{ Y_{ij},\Delta_{ij},\bfX_{ij}\}_{i=1,\cdots,n;j=1,\cdots,m_i}\right]\\
&= N^{-1}\sum_{i=1}^n \bfomega_i(\bfalpha_0)\bfomega_i(\bfalpha_0)^T=\Sigma.
\end{split}
\end{equation*}
It follows that given observed data, the conditional distribution of $\sqrt{N}(\hatalpha^{*} - \hatalpha_N)$ is asymptotically equivalent to the unconditional distribution of $\sqrt{N}(\hatalpha_N-\bfalpha_0)$, thereby justifying the perturbation-based covariance estimation procedure.

\end{proof}

\section{Additional Numerical Results}

\textbf{Scenario 5.} We consider a Frank copula model for the error terms, keeping the scheme for data generation the same as in Scenario 2. 

\textbf{Scenario 6.} Unlike the exchangeable dependence considered in Scenarios 1-3 and the new Scenario 5, we adopt an AR(1) dependence structure in this scenario. Particularly, the error terms satisfy $\exp(\epsilon_{ij})\sim Exp(1)$ marginally and are jointly modeled by a Guassian copula with AR(1) correlation of $0.7^{|j-k|}$ between $\epsilon_{ij}$ and $\epsilon_{ik}$ for $j\ne k$. All the other setups remain the same as in Scenario 2.

\textbf{Scenario 7.} A cluster-level covariate $x_{ij} = x_i$ is considered, where $x_{i}$ is independently generated from a Bernoulli$(0.5)$. The error terms $(\epsilon_{i1},\cdots,\epsilon_{im})$ follow a multivariate normal distribution with mean zero and covariance matrix $0.64\Sigma_{\epsilon}$, where the $(j,k)$-th entry of $\Sigma_{\epsilon}$ is $0.7^{|j-k|}$. The failure time outcomes $T_{ij}$ are generated from the AFT model %follows 
$\log T_{ij} = \beta_0+\beta_1x_i+\epsilon_{ij}$ with coefficients $\beta_0=0.5$ and $\beta_1 = 1$. In this setting, parameters in the QRL model (3) are obtained by 
\begin{equation*}
\alpha_0(\tau,t_0) = \left\{\begin{array}{cc}
   \beta_0+\Phi_{0.8}^{-1}(\tau),  &  t_0=0,\\
    \log\Big\{ -t_0+\exp\left[ \beta_0+\Phi_{0.8}^{-1}\left\{ (1-\tau)\Phi_{0.8}(\log t_0-\beta_0)+\tau\right\}\right] \Big\}, &  t_0\neq 0,
\end{array}\right.
\end{equation*}
\begin{equation*}
\alpha_1(\tau,t_0) = \left\{\begin{array}{cc}
   \beta_1,  &  t_0=0,\\
    \log\left\{ \frac{-t_0+\exp\left[ \beta_0+\beta_1+\Phi_{0.8}^{-1}\left\{ (1-\tau)\Phi_{0.8}(\log t_0-\beta_0-\beta_1)+\tau\right\}\right] }{-t_0+\exp\left[ \beta_0+\Phi_{0.8}^{-1}\left\{ (1-\tau)\Phi_{0.8}(\log t_0-\beta_0)+\tau\right\}\right] }\right\}, &  t_0\neq 0,
\end{array}\right.
\end{equation*}
where $\Phi_{0.8}(\cdot)$ is the cumulative distribution function of a zero-mean normal variable with a standard deviation of 0.8.

\textbf{Scenario 8.} The multivariate failure times is generated from the model: $\log T_{ij} = \beta_0+\beta_1x_{1i} + \beta_2x_{2i} + \beta_3x_{3ij}+\epsilon_{ij}$, where $x_{1i}\sim \text{Bernoulli}(0.5)$, $x_{2i}\sim\text{Uniform}[0,1]$, $x_{3ij}\sim N(0,1)$, $\beta_0 =\beta_1= 0.6$, $\beta_2 = 0.8$, $\beta_3 = 0.4$. The assumption for %generation of 
the error terms is the same as in Scenario 2. Under this setup, $\alpha_0(\tau,t_0) = \log[-\lambda^{-1}\log (1-\tau)]+\beta_0$, $\alpha_1(\tau,t_0) = \beta_1$, $\alpha_2(\tau,t_0)=\beta_2$ and $\alpha_3(\tau,t_0) = \beta_3$.

Estimation results under Scenarios 5-7 are summarized in Table \ref{table: sim5},  which consistently highlight the outperformance of the proposed marginal method in addressing diverse dependence structures for multivariate failure times even based on an independent working model. These promising findings further exhibit a degree of robustness of the method across different types of copula. Simulation results for Scenario 8, summarized in Table \ref{table: sim8}, 
also demonstrate that the overall performance of the proposed estimators is promising compared to the IFR estimator. Particularly, its superiority for the coefficients associated with cluster-level covariates $x_{1i}$ and $x_{2i}$ is greater than for the individual-level covariate $x_{3ij}$.

\begin{table}[h]
\footnotesize
\tabcolsep=5pt
\caption{Estimation results based on 500 replicates for quantile level $\tau=0.5$ under Scenario 1 with Kendall's tau=0.} 
\label{table: sim1-1}
\begin{center}
\begin{tabular}{llllllrlllc}
\hline\hline
\multicolumn{3}{c}{}&\multicolumn{3}{c}{$\alpha_0(0.5,t_0)$}&\multicolumn{1}{c}{}&\multicolumn{3}{c}{$\alpha_1(0.5,t_0)$}&\multicolumn{1}{c}{runtime}\tabularnewline
\cline{4-6}\cline{8-10}
$(n,m)$&&&$t_0=0$&$t_0=1$&$t_0=2$&&$t_0=0$&$t_0=1$&$t_0=2$&(s)\tabularnewline
\hline
(200,3)&bias&&-0.013&-0.011&-0.015&&0.026&0.024&0.028\tabularnewline
&MCSD&&0.13&0.147&0.17&&0.238&0.263&0.29\tabularnewline
&ASE&IFR&0.129&0.148&0.175&&0.231&0.261&0.300\tabularnewline
&&FR&0.132&0.15&0.176&&0.236&0.267&0.305&5.523\tabularnewline
&&CFS&0.128&0.156&0.191&&0.23&0.273&0.327&0.285\tabularnewline
&&RBS&0.126&0.143&0.165&&0.224&0.251&0.284&3.665\tabularnewline
&CP&IFR&0.946&0.946&0.971&&0.952&0.946&0.958\tabularnewline
&&FR&0.96&0.95&0.958&&0.95&0.962&0.958\tabularnewline
&&CFS&0.946&0.954&0.978&&0.952&0.954&0.972\tabularnewline
&&RBS&0.944&0.94&0.936&&0.936&0.942&0.944\tabularnewline
\cline{1-11}
(500,3)&bias&&-0.001&-0.003&-0.01&&0.006&0.01&0.015\tabularnewline
&MCSD&&0.081&0.092&0.112&&0.148&0.164&0.191\tabularnewline
&ASE&IFR&0.082&0.094&0.109&&0.148&0.166&0.188\tabularnewline
&&FR&0.082&0.092&0.107&&0.147&0.163&0.186&10.104\tabularnewline
&&CFS&0.081&0.098&0.119&&0.146&0.173&0.206&0.683\tabularnewline
&&RBS&0.08&0.089&0.103&&0.142&0.157&0.178&6.714\tabularnewline
&CP&IFR&0.952&0.958&0.959&&0.948&0.938&0.943\tabularnewline
&&FR&0.96&0.954&0.934&&0.95&0.956&0.946\tabularnewline
&&CFS&0.952&0.964&0.968&&0.946&0.946&0.97\tabularnewline
&&RBS&0.95&0.946&0.922&&0.944&0.946&0.938\tabularnewline
\cline{1-11}
(200,10)&bias&&-0.004&-0.006&-0.004&&0.009&0.014&0.011\tabularnewline
&MCSD&&0.071&0.079&0.092&&0.127&0.136&0.155\tabularnewline
&ASE&IFR&0.071&0.082&0.092&&0.128&0.145&0.161\tabularnewline
&&FR&0.071&0.081&0.091&&0.128&0.143&0.159&11.762\tabularnewline
&&CFS&0.07&0.085&0.101&&0.127&0.149&0.176&1.207\tabularnewline
&&RBS&0.069&0.079&0.089&&0.124&0.139&0.153&6.511\tabularnewline
&CP&IFR&0.952&0.952&0.962&&0.964&0.968&0.964\tabularnewline
&&FR&0.94&0.958&0.95&&0.95&0.96&0.952\tabularnewline
&&CFS&0.948&0.962&0.972&&0.964&0.968&0.976\tabularnewline
&&RBS&0.938&0.944&0.942&&0.946&0.954&0.946\tabularnewline
\hline
%0.25 & (200,3)&bias&&-0.001&0.006&0.003&&0.004&-0.01&-0.004\tabularnewline
%&&MCSD&&0.137&0.15&0.168&&0.237&0.261&0.288\tabularnewline
%&&ASE&FR&0.137&0.152&0.177&&0.235&0.266&0.306&5.214\tabularnewline
%&&&CFS&0.136&0.158&0.19&&0.232&0.273&0.326&0.245\tabularnewline
%&&&RBS&0.131&0.145&0.165&&0.221&0.249&0.283&3.712\tabularnewline
%&&CP&FR&0.944&0.94&0.968&&0.954&0.948&0.96\tabularnewline
%&&&CFS&0.942&0.962&0.978&&0.946&0.954&0.976\tabularnewline
%&&&RBS&0.928&0.934&0.954&&0.936&0.924&0.948\tabularnewline
%\cline{2-12}
%& (500,3)&bias&&0.001&-0.001&-0.003&&-0.003&-0.002&0.001\tabularnewline
%&&MCSD&&0.084&0.095&0.099&&0.142&0.159&0.172\tabularnewline
%&&ASE&FR&0.086&0.095&0.11&&0.146&0.166&0.191&10.675\tabularnewline
%&&&CFS&0.086&0.099&0.119&&0.146&0.172&0.204&0.816\tabularnewline
%&&&RBS&0.084&0.092&0.106&&0.141&0.16&0.183&6.644\tabularnewline
%&&CP&FR&0.952&0.94&0.968&&0.946&0.96&0.966\tabularnewline
%&&&CFS&0.964&0.954&0.966&&0.944&0.962&0.96\tabularnewline
%&&&RBS&0.946&0.936&0.964&&0.94&0.956&0.966\tabularnewline
%\cline{2-12}
%&(200,10)&bias&&-0.002&-0.002&0&&0.001&0.005&0.004\tabularnewline
%&&MCSD&&0.083&0.087&0.093&&0.133&0.153&0.161\tabularnewline
%&&ASE&FR&0.086&0.087&0.096&&0.128&0.144&0.162&11.603\tabularnewline
%&&&CFS&0.085&0.091&0.106&&0.125&0.15&0.179&1.066\tabularnewline
%&&&RBS&0.083&0.085&0.093&&0.124&0.14&0.156&6.417\tabularnewline
%&&CP&FR&0.952&0.954&0.954&&0.938&0.936&0.946\tabularnewline
%&&&CFS&0.954&0.966&0.98&&0.944&0.958&0.972\tabularnewline
%&&&RBS&0.948&0.95&0.95&&0.926&0.934&0.936\tabularnewline
%\hline
\end{tabular}\end{center}
\end{table}

\begin{table}[htbp]
\footnotesize
\caption{Estimation results based on 500 replicates for quantile level $\tau=0.5$ under Scenario 1 with Kendall's tau=0.8.}
\label{table: sim1-3}
\begin{center}
\begin{tabular}{llllllrlllc}
\hline\hline
\multicolumn{3}{c}{}&\multicolumn{3}{c}{$\alpha_0(0.5,t_0)$}&\multicolumn{1}{c}{}&\multicolumn{3}{c}{$\alpha_1(0.5,t_0)$}&\multicolumn{1}{c}{runtime}\tabularnewline
\cline{4-6}\cline{8-10}
$(n,m)$&&&$t_0=0$&$t_0=1$&$t_0=2$&&$t_0=0$&$t_0=1$&$t_0=2$&(s)\tabularnewline
\hline
(200,3)&bias&&-0.011&-0.007&-0.023&&0.022&0.014&0.017&\tabularnewline
&MCSD&&0.145&0.154&0.177&&0.229&0.253&0.28&\tabularnewline
&ASE&IFR&0.127&0.147&0.171&&0.23&0.259&0.295&\tabularnewline
&&FR&0.151&0.171&0.16&&0.237&0.27&0.255&5.905\tabularnewline
&&CFS&0.15&0.178&0.176&&0.232&0.279&0.275&0.288\tabularnewline
&&RBS&0.143&0.162&0.15&&0.223&0.251&0.236&3.603\tabularnewline
&CP&IFR&0.914&0.95&0.935&&0.94&0.956&0.957&\tabularnewline
&&FR&0.958&0.976&0.921&&0.954&0.964&0.931&\tabularnewline
&&CFS&0.956&0.978&0.947&&0.942&0.974&0.945&\tabularnewline
&&RBS&0.954&0.96&0.919&&0.934&0.946&0.911&\tabularnewline
\cline{1-11}
(500,3)&bias&&0.003&0.006&0.007&&0&-0.006&-0.005&\tabularnewline
&MCSD&&0.091&0.113&0.124&&0.143&0.174&0.198&\tabularnewline
&ASE&IFR&0.082&0.094&0.107&&0.147&0.166&0.187&\tabularnewline
&&FR&0.096&0.108&0.113&&0.149&0.17&0.181&10.822\tabularnewline
&&CFS&0.095&0.113&0.123&&0.147&0.176&0.196&0.636\tabularnewline
&&RBS&0.093&0.104&0.108&&0.143&0.163&0.173&5.685\tabularnewline
&CP&IFR&0.926&0.892&0.908&&0.954&0.926&0.945&\tabularnewline
&&FR&0.96&0.934&0.923&&0.956&0.926&0.936&\tabularnewline
&&CFS&0.958&0.948&0.947&&0.952&0.936&0.949&\tabularnewline
&&RBS&0.952&0.928&0.915&&0.944&0.924&0.919&\tabularnewline
\cline{1-11}
(200,10)&bias&&0.001&0.001&-0.005&&0.003&0.004&-0.005&\tabularnewline
&MCSD&&0.112&0.124&0.131&&0.122&0.144&0.178&\tabularnewline
&ASE&IFR&0.069&0.081&0.092&&0.125&0.142&0.16&\tabularnewline
&&FR&0.114&0.126&0.121&&0.13&0.155&0.162&13.032\tabularnewline
&&CFS&0.113&0.131&0.134&&0.127&0.159&0.176&0.765\tabularnewline
&&RBS&0.108&0.12&0.116&&0.127&0.15&0.156&6.002\tabularnewline
&CP&IFR&0.776&0.8&0.822&&0.948&0.938&0.921&\tabularnewline
&&FR&0.954&0.944&0.932&&0.956&0.948&0.935&\tabularnewline
&&CFS&0.954&0.958&0.955&&0.952&0.956&0.955&\tabularnewline
&&RBS&0.944&0.93&0.926&&0.952&0.944&0.917&\tabularnewline
\hline
\end{tabular}\end{center}
\end{table}

\begin{table}[htbp]
\footnotesize
\tabcolsep=3pt
\caption{Estimation results based on 500 replicates for quantile level $\tau=0.5$ under Scenarios 5-7 ($n=200,m=10$).} 
\label{table: sim5}
%simdata=8
\begin{center}
\begin{tabular}{llllllrlllc}
\hline\hline
\multicolumn{3}{c}{}&\multicolumn{3}{c}{$\alpha_0(0.5,t_0)$}&\multicolumn{1}{c}{}&\multicolumn{3}{c}{$\alpha_1(0.5,t_0)$}&\multicolumn{1}{c}{runtime}\tabularnewline
\cline{4-6}\cline{8-10}
Scenario&&&$t_0=0$&$t_0=1$&$t_0=2$&&$t_0=0$&$t_0=1$&$t_0=2$&(s)\tabularnewline
\hline
5&bias&&-0.002&-0.006&-0.006&&0.001&0.011&0.012&\tabularnewline
&MCSD&&0.16&0.159&0.163&&0.286&0.281&0.284&\tabularnewline
&ASE&IFR&0.071&0.081&0.093&&0.129&0.144&0.163&32.295\tabularnewline
&&FR&0.16&0.157&0.162&&0.282&0.277&0.284&27.902\tabularnewline
&&CFS&0.161&0.165&0.18&&0.282&0.292&0.314&1.234\tabularnewline
&&RBS&0.155&0.151&0.156&&0.27&0.265&0.273&11.775\tabularnewline
&CP&IFR&0.6&0.684&0.748&&0.598&0.682&0.75&\tabularnewline
&&FR&0.956&0.954&0.95&&0.954&0.952&0.946&\tabularnewline
&&CFS&0.956&0.96&0.964&&0.954&0.966&0.97&\tabularnewline
&&RBS&0.954&0.94&0.936&&0.952&0.942&0.94&\tabularnewline
\hline
6&bias&&0.002&0&-0.007&&-0.006&-0.007&0.001&\tabularnewline
&MCSD&&0.113&0.106&0.121&&0.203&0.194&0.213&\tabularnewline
&ASE&IFR&0.069&0.081&0.099&&0.122&0.141&0.167&31.735\tabularnewline
&&FR&0.116&0.113&0.125&&0.204&0.199&0.215&36.391\tabularnewline
&&CFS&0.116&0.12&0.139&&0.203&0.21&0.238&1.888\tabularnewline
&&RBS&0.113&0.111&0.122&&0.197&0.194&0.209&19.061\tabularnewline
&CP&IFR&0.754&0.872&0.898&&0.742&0.86&0.866&\tabularnewline
&&FR&0.97&0.966&0.968&&0.96&0.95&0.96&\tabularnewline
&&CFS&0.968&0.97&0.98&&0.956&0.966&0.978&\tabularnewline
&&RBS&0.958&0.96&0.964&&0.948&0.946&0.942&\tabularnewline
\hline
7&bias&&-0.006&-0.012&-0.009&&0.008&0.013&0.011&\tabularnewline
&MCSD&&0.059&0.086&0.099&&0.082&0.11&0.132&\tabularnewline
&SE&IFR&0.033&0.057&0.083&&0.049&0.073&0.101&29.449\tabularnewline
&&FR&0.057&0.08&0.1&&0.082&0.108&0.129&28.179\tabularnewline
&&CFS&0.056&0.085&0.112&&0.082&0.114&0.143&1.384\tabularnewline
&&RBS&0.056&0.078&0.097&&0.08&0.106&0.125&14.203\tabularnewline
&CP&IFR&0.744&0.798&0.896&&0.764&0.816&0.874&\tabularnewline
&&FR&0.936&0.942&0.948&&0.952&0.958&0.946&\tabularnewline
&&CFS&0.936&0.956&0.966&&0.952&0.966&0.976&\tabularnewline
&&RBS&0.932&0.924&0.942&&0.95&0.95&0.94&\tabularnewline
\hline
\end{tabular}\end{center}
\end{table}

\begin{table}[htbp]
\tiny
\tabcolsep=3pt
\caption{Estimation results based on 500 replicates for quantile level $\tau=0.5$ under Scenario 8.} 
\label{table: sim8}
%simdata=7
\begin{center}
\begin{tabular}{llllllrlllrlllrlll}
\hline\hline
\multicolumn{3}{c}{}&\multicolumn{3}{c}{$\alpha_0(0.5,t_0)$}&\multicolumn{1}{c}{}&\multicolumn{3}{c}{$\alpha_1(0.5,t_0)$}&\multicolumn{1}{c}{}&\multicolumn{3}{c}{$\alpha_2(0.5,t_0)$}&\multicolumn{1}{c}{}&\multicolumn{3}{c}{$\alpha_3(0.5,t_0)$}\tabularnewline
\cline{4-6}\cline{8-10}\cline{12-14}\cline{16-18}
$(n,m)$&&&$t_0=0$&$t_0=1$&$t_0=2$&&$t_0=0$&$t_0=1$&$t_0=2$&&$t_0=0$&$t_0=1$&$t_0=2$&&$t_0=0$&$t_0=1$&$t_0=2$\tabularnewline
\hline
(200,3)&bias&&-0.015&-0.006&-0.012&&0.005&-0.003&0.008&&0.016&0.014&0.007&&0.004&0.005&0.001\tabularnewline
&MCSD&&0.195&0.206&0.243&&0.17&0.179&0.204&&0.296&0.307&0.351&&0.066&0.084&0.099\tabularnewline
&ASE&IFR&0.151&0.18&0.214&&0.138&0.159&0.184&&0.242&0.277&0.323&&0.07&0.081&0.094\tabularnewline
&&FR&0.203&0.211&0.233&&0.184&0.189&0.204&&0.322&0.329&0.357&&0.07&0.082&0.095\tabularnewline
&&CFS&0.196&0.212&0.245&&0.179&0.192&0.217&&0.309&0.333&0.379&&0.067&0.082&0.1\tabularnewline
&&RBS&0.191&0.199&0.219&&0.174&0.177&0.191&&0.3&0.308&0.336&&0.069&0.081&0.095\tabularnewline
&CP&IFR&0.882&0.918&0.924&&0.882&0.926&0.928&&0.884&0.928&0.924&&0.956&0.948&0.954\tabularnewline
&&FR&0.962&0.954&0.95&&0.968&0.96&0.954&&0.962&0.966&0.956&&0.958&0.952&0.958\tabularnewline
&&CFS&0.952&0.956&0.96&&0.964&0.96&0.97&&0.954&0.966&0.97&&0.944&0.948&0.966\tabularnewline
&&RBS&0.942&0.942&0.938&&0.952&0.95&0.934&&0.952&0.942&0.94&&0.95&0.946&0.958\tabularnewline
\hline
(500,3)&bias&&-0.007&-0.006&-0.004&&0.005&0&-0.006&&0.002&0.009&0.016&&0.003&0.005&0.002\tabularnewline
&MCSD&&0.127&0.13&0.144&&0.119&0.123&0.126&&0.195&0.193&0.21&&0.044&0.052&0.062\tabularnewline
&ASE&IFR&0.092&0.111&0.133&&0.085&0.099&0.115&&0.148&0.173&0.2&&0.044&0.051&0.059\tabularnewline
&&FR&0.126&0.132&0.146&&0.115&0.119&0.129&&0.199&0.207&0.224&&0.044&0.052&0.059\tabularnewline
&&CFS&0.124&0.135&0.157&&0.113&0.122&0.139&&0.195&0.212&0.241&&0.043&0.053&0.065\tabularnewline
&&RBS&0.12&0.126&0.141&&0.11&0.115&0.124&&0.189&0.198&0.214&&0.043&0.051&0.06\tabularnewline
&CP&IFR&0.85&0.92&0.928&&0.842&0.894&0.924&&0.86&0.934&0.938&&0.936&0.94&0.932\tabularnewline
&&FR&0.944&0.954&0.952&&0.94&0.934&0.958&&0.962&0.968&0.97&&0.936&0.946&0.932\tabularnewline
&&CFS&0.942&0.958&0.966&&0.932&0.95&0.972&&0.962&0.972&0.978&&0.934&0.954&0.946\tabularnewline
&&RBS&0.936&0.948&0.94&&0.924&0.932&0.938&&0.958&0.96&0.958&&0.934&0.94&0.932\tabularnewline
\hline
(200,10)&bias&&-0.006&-0.01&-0.011&&-0.001&-0.002&-0.003&&-0.005&0.004&0.01&&0.001&-0.001&0.001\tabularnewline
&MCSD&&0.175&0.152&0.15&&0.161&0.143&0.142&&0.28&0.252&0.244&&0.037&0.043&0.051\tabularnewline
&ASE&IFR&0.08&0.097&0.115&&0.074&0.086&0.099&&0.128&0.15&0.174&&0.037&0.044&0.051\tabularnewline
&&FR&0.175&0.161&0.16&&0.159&0.147&0.144&&0.274&0.256&0.253&&0.038&0.046&0.053\tabularnewline
&&CFS&0.172&0.166&0.174&&0.157&0.153&0.158&&0.269&0.265&0.276&&0.037&0.047&0.058\tabularnewline
&&RBS&0.165&0.155&0.155&&0.151&0.142&0.139&&0.261&0.244&0.244&&0.038&0.046&0.054\tabularnewline
&CP&IFR&0.636&0.81&0.862&&0.642&0.752&0.836&&0.636&0.754&0.834&&0.94&0.954&0.952\tabularnewline
&&FR&0.944&0.958&0.978&&0.96&0.962&0.95&&0.944&0.962&0.97&&0.944&0.958&0.958\tabularnewline
&&CFS&0.938&0.97&0.984&&0.956&0.972&0.974&&0.94&0.968&0.986&&0.94&0.958&0.976\tabularnewline
&&RBS&0.932&0.95&0.968&&0.946&0.962&0.942&&0.932&0.956&0.96&&0.944&0.958&0.96\tabularnewline
\hline
\end{tabular}\end{center}
\end{table}

\begin{figure}[ht]
\caption{Estimation results with different quantile levels and residual time points for the Framingham heart data.}
    \centering
    \includegraphics{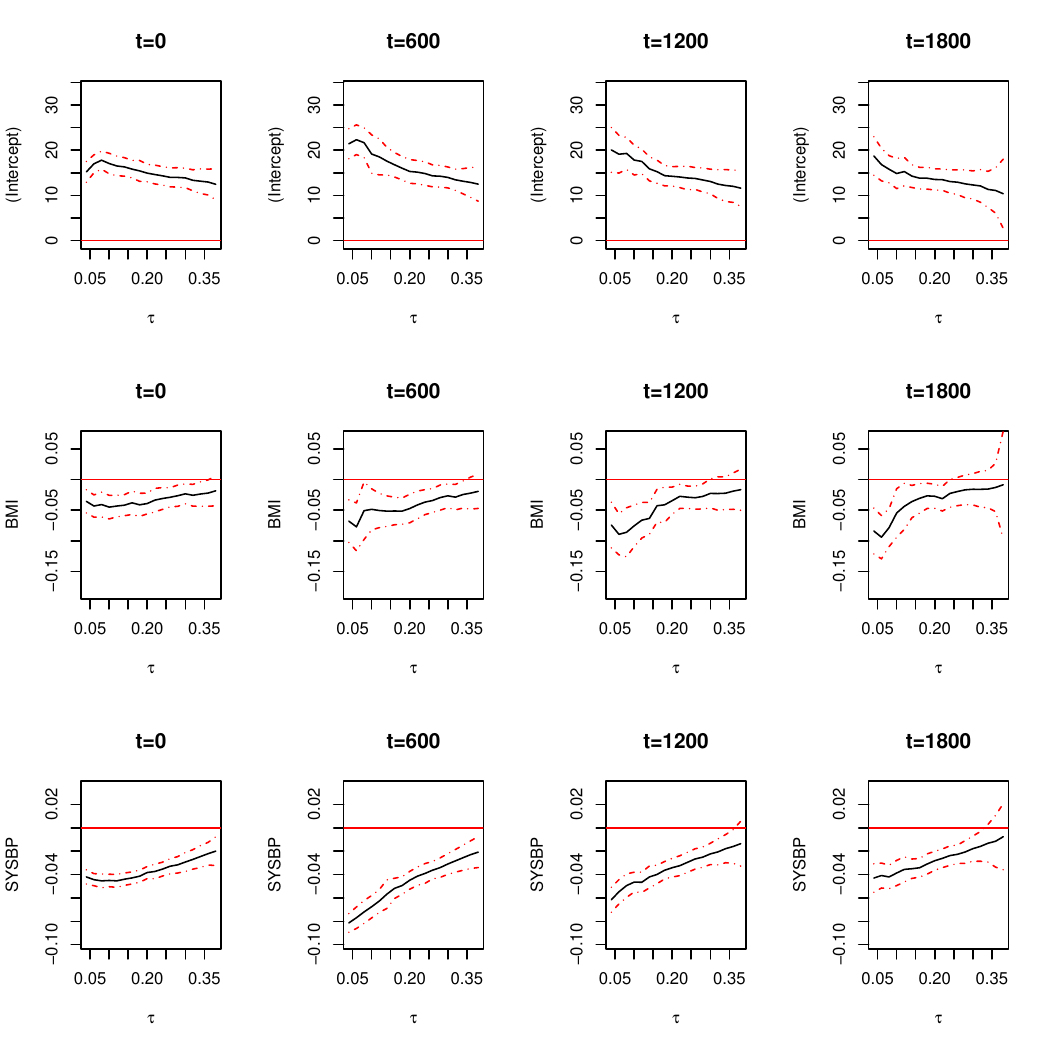}
\end{figure}

\begin{figure}
\caption{Estimation results %of the framingham study for 
with different quantile levels and residual time points for the Framingham heart data.}
    \centering
    \includegraphics{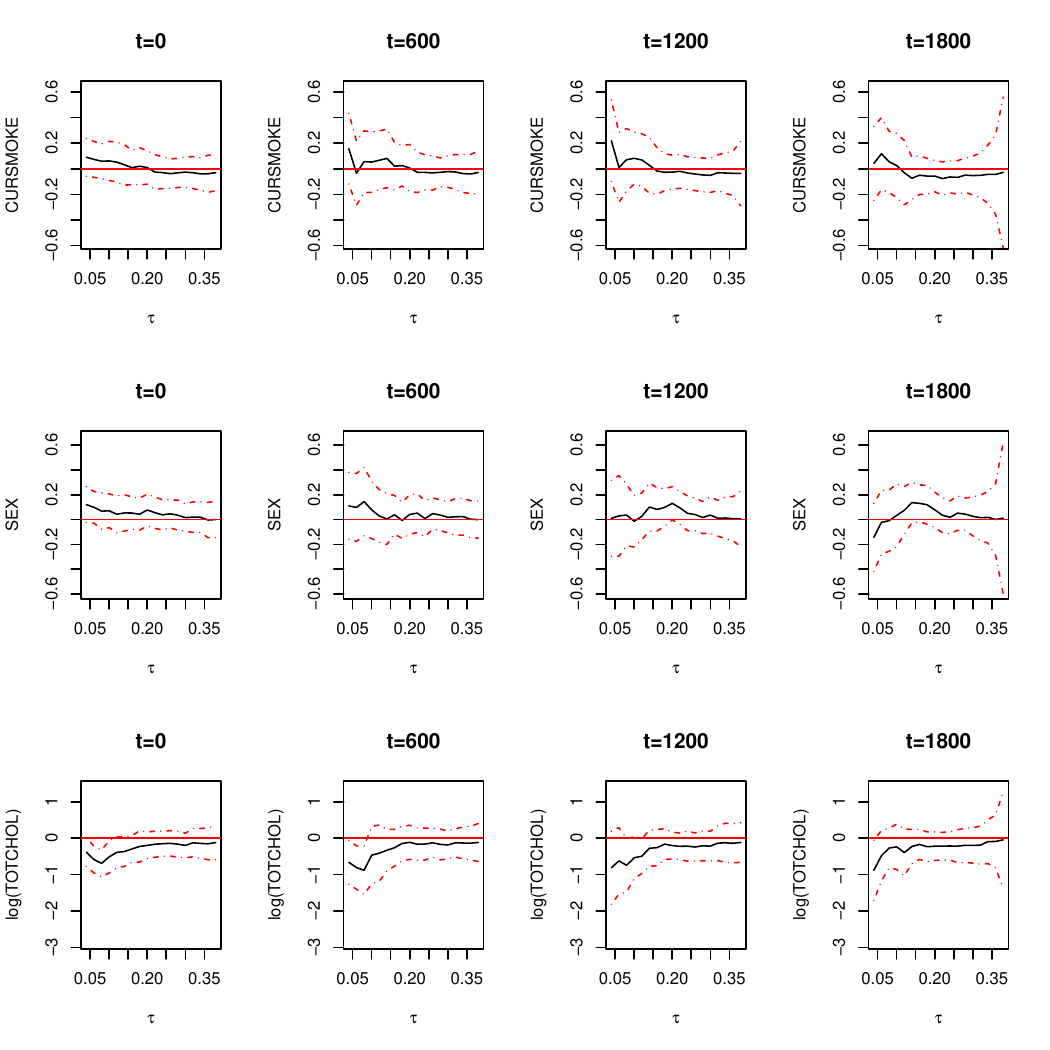}
\end{figure}

\end{document}